\documentclass[aps,twocolumn,amsmath,amssymb,floatfix,prl,superscriptaddress]{revtex4-1}

\usepackage{graphicx}
\usepackage{dcolumn}
\usepackage{bm}
\usepackage{latexsym}
\usepackage{color}
\usepackage{xcolor}
\usepackage[colorlinks=true,linkcolor=blue,citecolor=blue]{hyperref}

\begin{document}

\title{Physics of Weibel-mediated relativistic collisionless shocks}

\author{Martin Lemoine}
\affiliation{Institut d'Astrophysique de Paris, CNRS -- Sorbonne Universit\'e, 98 bis boulevard Arago, F-75014 Paris}
\author{Laurent Gremillet}
\affiliation{CEA, DAM, DIF, F-91297 Arpajon, France}
\author{Guy Pelletier} ,
\affiliation{Universit\'e Grenoble Alpes, CNRS-INSU, Institut de Plan\'etologie et d'Astrophysique de Grenoble (IPAG), F-38041 Grenoble, France}
\author{Arno Vanthieghem}
\affiliation{Institut d'Astrophysique de Paris, CNRS -- Sorbonne Universit\'e, 98 bis boulevard Arago, F-75014 Paris}
\affiliation{Sorbonne Universit\'e, Institut Lagrange de Paris (ILP),
98 bis bvd Arago, F-75014 Paris, France}

\date{\today}

\begin{abstract} 
We develop a comprehensive theoretical model of relativistic collisionless pair shocks mediated by the current filamentation instability. We notably characterize the noninertial frame in which this instability is of a mostly magnetic nature, and describe at a microscopic level the deceleration and heating of the incoming background plasma through its collisionless interaction with the electromagnetic turbulence. Our model compares well to large-scale 2D3V PIC simulations, and provides an important touchstone for the phenomenology of such plasma systems.
\end{abstract}

\maketitle

\noindent
\emph{Introduction--} 
Though of mundane occurrence in space plasmas, collisionless shock waves represent outstanding phenomena owing to their inherent complexity and many astrophysical repercussions~\cite{2009A&ARv..17..409T, *2011A&ARv..19...42B}. Mediated by collective electromagnetic interactions, whose nature and origin form the focus of active investigations~\cite{2016RPPh...79d6901M}, they seemingly  accelerate charged particles to high energies in a broad variety of sources~\cite{1987PhR...154....1B}, giving rise to a rich phenomenology at the core of high-energy and multi-messenger astrophysics. The electromagnetic counterpart of the gravitational wave event GW170817 is thus interpreted as the synchrotron radiation of electrons energized at the unmagnetized, relativistic shock wave triggered by the neutron star coalescence~\cite{PhysRevLett.119.161101, *2041-8205-848-2-L12}. In parallel, collisionless shocks have become central topics in high-power laser-plasma experiments, which might well generate and study such structures in the near future~\cite{PhysRevLett.111.225002, *Huntington_NP_11_173_2015, *PhysRevLett.118.185003}.

In the absence of a significant background magnetic field, the physics of the shock is governed by an electromagnetic microturbulence driven by a current filamentation instability (CFI), as predicted~\cite{1963JNuE....5...43M, *1999ApJ...526..697M, *2006ApJ...647.1250L, *2007A&A...475....1A, *2007A&A...475...19A}, and as observed in {\it ab initio} simulations~\cite{2007ApJ...668..974K, *2008ApJ...673L..39S, *2008ApJ...681L..93K, *2008ApJ...682L...5S,*2009ApJ...698L..10N, *2009ApJ...695L.189M, *2011ApJ...726...75S, *2011ApJ...739L..42H}. This microturbulence dissipates the ordered kinetic energy of the unshocked plasma, just as it governs the acceleration of particles to suprathermal energies. The latter, in turn, induce electromagnetic instabilities in the upstream region~\cite{2008PhRvL.100t5008B, *2010PhPl...17l0501B, *2010PhRvE..81c6402B, *2010MNRAS.402..321L, *2011ApJ...736..157R, *2011MNRAS.417.1148L, *2012ApJ...744..182S}, thereby ensuring the self-sustained and (quasi-) stationary nature of the shock. Recent theoretical models have discussed the formation of the shock~\cite{2013PhPl...20d2102B, *2014PhPl...21g2301B}, the structure of the microturbulence~\cite{2007ApJ...655..375K, *2009ApJ...696.2269M}, or the early-time shock transition in the sub-relativistic regime \cite{2016PhRvL.117f5001R, *2017PhPl...24d1409R}, but a detailed microphysical picture of well-formed shocks remains missing.

In this Letter, we present a comprehensive theoretical model for unmagnetized, relativistic collisionless shock waves, such as those expected at the boundary of relativistic astrophysical jets. Specifically, we provide a microphysical description of the deceleration and nonadiabatic heating of the background plasma in the shock precursor, and of the dynamics of the microturbulence and suprathermal particles. Our model relies on the observation that there exists a noninertial frame (hereafter ``Weibel frame'') in which the microturbulence is essentially magnetostatic. Introducing such a frame allows one to derive proper transport equations for the background and suprathermal particles. Our arguments are shown to agree with dedicated high-resolution, large-scale particle-in-cell (PIC) simulations conducted using the code \textsc{calder}~\cite{2003NucFu..43..629L} in a 2D3V geometry (2D in configuration space, 3D in momentum space) \cite{L1_supp}. We restrict ourselves to the case of a shock propagating in an electron-positron plasma, but discuss how the results can be generalized to electron-ion plasmas. 

\begin{figure}
\includegraphics[width=0.45\textwidth]{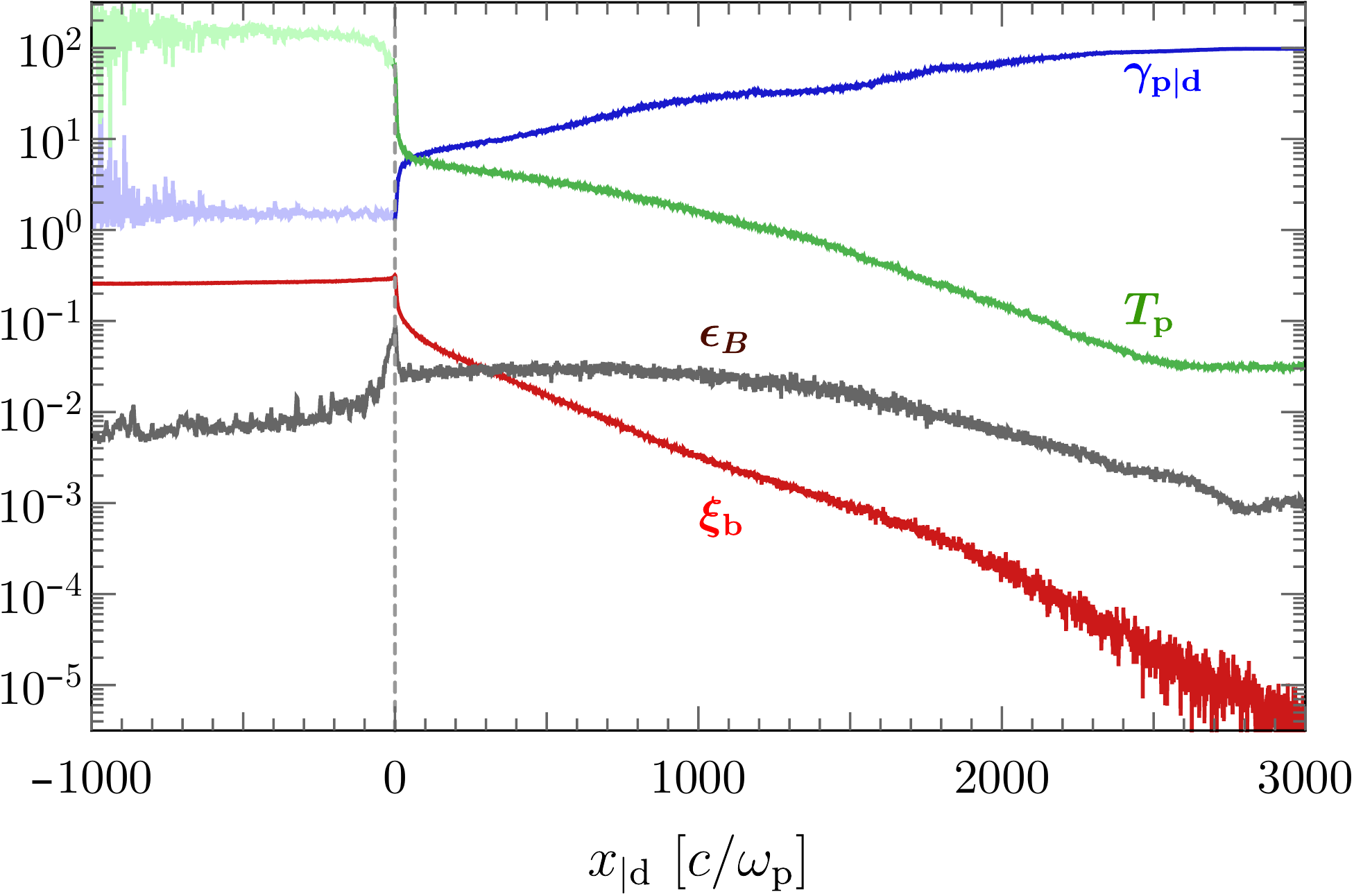}
 \caption{Downstream/simulation frame 1D spatial profiles of the background plasma Lorentz factor $\gamma_{\rm p\vert\rm d}$, of its proper temperature $T_{\rm p}$ (units $m_e c^2/k_{\rm B}$), of the suprathermal beam pressure $\xi_{\rm b}$ and of the microturbulence energy density $\epsilon_B$ for a 2D3V PIC simulation of initial Lorentz factor $\gamma_\infty = 173$ ($\gamma_{\infty\vert\rm d} = 100$ in the simulation frame). Data are light colored in regions where they cannot be measured accurately.
  \label{fig:prof_ga100} }
\end{figure}

We describe the 1D profile of a formed shock, assumed stationary in the shock front rest frame $\mathcal R_{\rm s}$. The precursor is defined as the region where the background plasma coexists with a population of suprathermal particles, characterized by their pressure $\xi_{\rm b}$ normalized to the incoming momentum flux density $F_\infty \equiv \gamma_\infty^2\beta_\infty^2 n_\infty m_e c^2$ (with $\gamma_\infty$, $\beta_\infty$ and $n_\infty$ denoting, respectively, the Lorentz factor of the unshocked background plasma, its normalized velocity and its proper density), and a self-generated electromagnetic microturbulence of energy density $\epsilon_B$ (also in units of $F_\infty$)~\footnote{Quantities related to the background plasma (resp. suprathermal beam) are indexed with $_{\rm p}$ (resp. $_{\rm b}$). Also, subscripts $_{\vert\rm d}$ ($_{\vert\rm p}$) refer to quantities measured in the simulation (downstream) rest frame (resp. the background plasma rest frame).}. Both $\xi_{\rm b}$ and $\epsilon_B$ vary with the distance $x$ to the shock. Figure~\ref{fig:prof_ga100} plots their (transversely averaged) profiles extracted from a PIC simulation, in which the background plasma is injected with $\gamma_\infty = 173$ (\emph{i.e.}, $\gamma_{\infty\vert\rm d} = 100$ in the simulation frame, which coincides with the downstream rest frame) and proper temperature $T_{\rm p} = 10^{-2} m_e c^2/k_{\rm B}$. Distances are in units of $c/\omega_{\rm p} = c/\left(4\pi n_\infty e^2/m_e\right)^{1/2}$.

\noindent
\emph{The ``Weibel frame''--}
The leading micro-instability in driving the shock transition is the Weibel-type transverse CFI \cite{1963JNuE....5...43M, 2008PhRvL.100t5008B}, which fragments the background plasma into current filaments along the shock normal, surrounded by transverse electromagnetic fields $\boldsymbol{\delta}\mathbf{E}_{_{\boldsymbol{\perp}}}$ (radial) and $\boldsymbol{\delta}\mathbf{B}_{_{\boldsymbol{\perp}}}$ (toroidal). This instability is essentially magnetic, \emph{i.e.}, $\boldsymbol{\delta}\mathbf{B}_{_{\boldsymbol{\perp}}}^2-\boldsymbol{\delta}\mathbf{E}_{_{\boldsymbol{\perp}}}^2 > 0$. Along with $\boldsymbol{\delta}\mathbf{E}_{_{\boldsymbol{\perp}}}\cdot\boldsymbol{\delta}\mathbf{B}_{_{\boldsymbol{\perp}}} = 0$, this implies that, at a given point $x$, one can define a local reference frame, denoted $\mathcal R_{\rm w}$, where $\boldsymbol{\delta}\mathbf{E}_{_{\boldsymbol{\perp}}} = 0$. 
Far from the shock front, however, the transverse CFI might be superseded by electrostatic two-stream or oblique modes~\cite{2010PhRvE..81c6402B}, thus compromising the very existence of $\mathcal R_{\rm w}$. We indeed observe a finite $\delta E_x$, yet its energy density is well sub-dominant relative to that of $\delta B_{_\perp}$ in the near precursor. We therefore omit $\delta E_x$ for now, but we will comment on its possible role further on. Our model thus describes the turbulence as a collection of magnetostatic modes transverse to the flow in $\mathcal R_{\rm w}$.

\begin{figure}
\includegraphics[width=0.45\textwidth]{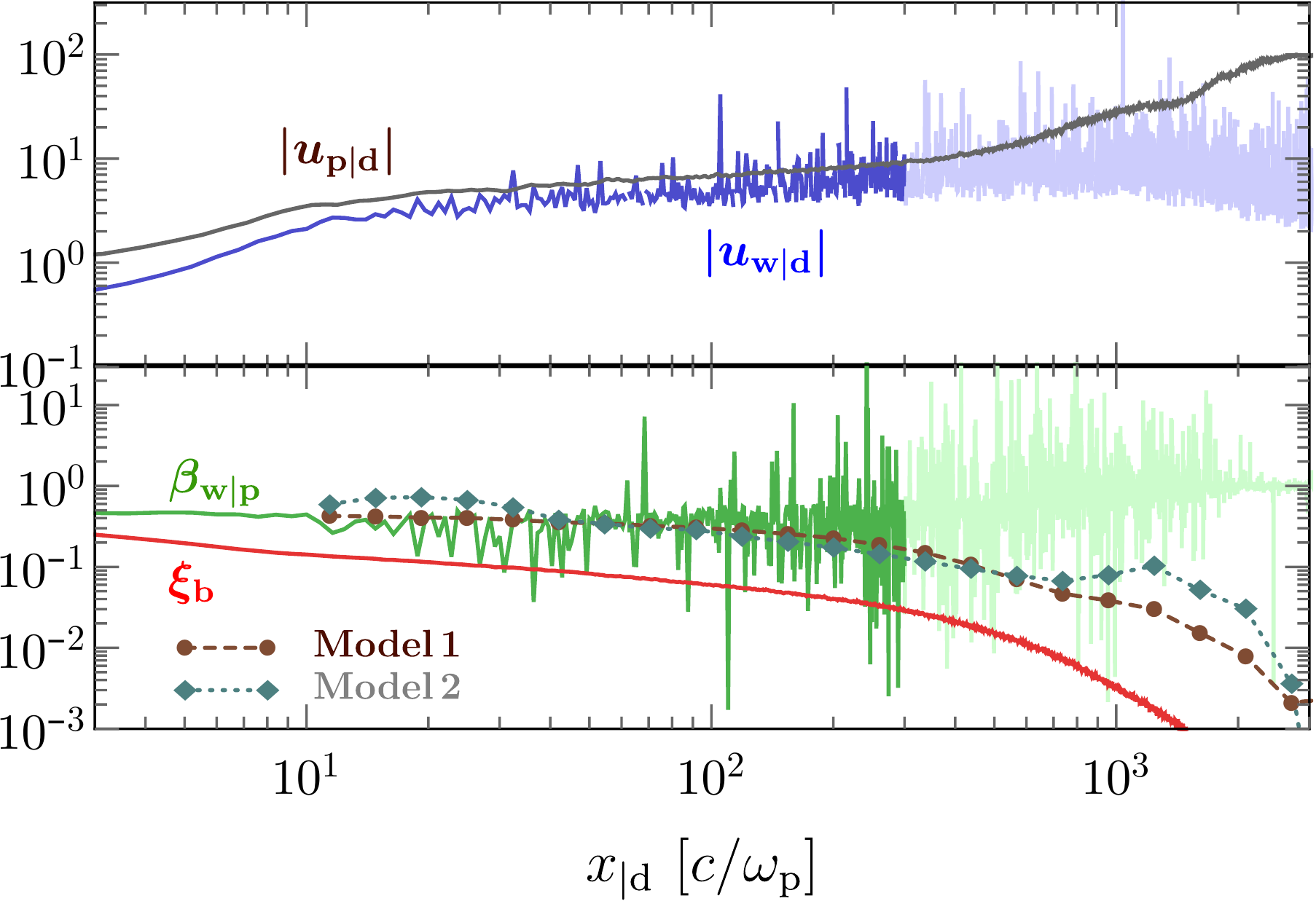}
 \caption{Top panel: $4$-velocities $\vert u_{\rm p\vert d}\vert$ and $\vert u_{\rm w\vert d}\vert$ measured in the PIC simulation with $\gamma_{\infty \vert\rm d} = 100$. Bottom panel: relative $3$-velocity $\beta_{\rm w\vert p}$ between $\mathcal R_{\rm w}$ and the background plasma compared to our two theoretical models, and suprathermal beam pressure $\xi_{\rm b}$. Data are light colored in regions  where they cannot be measured accurately: at $x \gtrsim 300 c/\omega_{\rm p}$, where $\vert\beta_{\rm w\vert d}\vert \simeq 1$, the estimate of $\beta_{\rm w\vert\rm p}$ carries a numerical error amplified by $\sim\gamma_{\rm w\vert d}^2$.
  \label{fig:bwp} }
\end{figure}

Figure~\ref{fig:bwp} displays the downstream-frame $4$-velocity $u_{\rm w\vert\rm d} = \gamma_{\rm w\vert\rm d}\beta_{\rm w\vert\rm d}$ of $\mathcal R_{\rm w}$, as extracted from the PIC simulation through the ratio $\langle\delta E_y^2\rangle^{1/2}/\langle\delta B_z^2\rangle^{1/2} = \beta_{\rm w\vert\rm d}$ (where averaging is done over the transverse dimension). 
That $\left\vert\beta_{\rm w\vert d}\right\vert<1$ confirms that $\mathcal R_{\rm w}$ is well defined, at least in the near precursor  $x \lesssim 10^3 c/\omega_{\rm p}$ where it can be
measured unambiguously, and where the shock transition mainly takes place. The spatial dependence of $u_{\rm w\vert d}$ indicates that $\mathcal R_{\rm w}$ is not globally inertial, which bears critical consequences for plasma heating, as explained below.

Obtaining a theoretical estimate of $\beta_{\rm w}$ turns out to be a nontrivial task. We determine this velocity through two approaches~\cite{pap1}: (i) we search for a frame, at each point along the precursor, where the fastest-growing CFI mode computed from the kinetic linear dispersion relation has a vanishing electrostatic component;
(ii) we search for a frame in which we can describe the nonlinear stage of the CFI as a locally stationary pressure equilibrium between the plasma, the beam and purely magnetic structures. Both approaches yield rather comparable estimates, $\beta_{\rm w\vert p} \propto \xi_{\rm b}$, with one important implication: $\mathcal R_{\rm w}$ moves at subrelativistic velocities relative to the background plasma, hence at relativistic velocities towards the shock front, with $\gamma_{\rm w} \lesssim \gamma_{\rm p}$. The magnitude of $\beta_{\rm w\vert p}$ proves to be a central element of our model.

Figure~\ref{fig:bwp} clearly illustrates these features: in PIC simulations, the background plasma moves slightly faster than $\mathcal R_{\rm w}$, and at $x\gtrsim100c/\omega_{\rm p}$, both $4$-velocities remain close to each other; the relative velocity $\beta_{\rm w\vert p}$ between the $\mathcal R_{\rm w}$ frame and the background plasma is always subrelativistic where it can be measured accurately; finally, our theoretical estimates of $\beta_{\rm w\vert p}$ agree well with the simulation data. Comparison of Figs.~\ref{fig:prof_ga100} and \ref{fig:bwp} also confirms that $\xi_{\rm b}$ provides a reasonable guide for the scaling of $\beta_{\rm w\vert p}$. That $\vert u_{\rm w\vert p}\vert < 1$ results from the large asymmetry between the background and suprathermal plasmas: in $\mathcal R_{\rm w}$ the latter forms a tenuous beam of large-inertia particles, which undergo small-angle scattering off the microturbulence; the former is comparatively dense and cold over most of the precursor, and its particles are mostly trapped in the magnetic filaments. 

\noindent
\emph{The deceleration of the background plasma--}
A nonvanishing $\xi_{\rm b}$ implies a nonvanishing $\beta_{\rm w\vert p}$, so that the frame $\mathcal R_{\rm w}$ never exactly coincides with the rest frame of the background plasma, which nevertheless keeps relaxing in $\mathcal R_{\rm w}$ through scattering. Hence, the finite pressure of the beam leads to the progressive deceleration of $\mathcal R_{\rm w}$, and, in turn, of the background plasma. This offers a view of how, at the kinetic level in $\mathcal R_{\rm w}$, momentum is transferred from the suprathermal beam to the background plasma. This explanation departs from the standard picture in which the CFI builds up a magnetized barrier \emph{in the shock rest frame}, which halts and isotropizes the incoming plasma particles~\cite{1963JNuE....5...43M, *1999ApJ...526..697M,  *2006ApJ...647.1250L, *2007A&A...475....1A, *2007A&A...475...19A}. As a matter of fact, if the scattering center frame were exactly static in the shock frame, Fermi acceleration would not occur.

\begin{figure}
\includegraphics[width=0.45\textwidth]{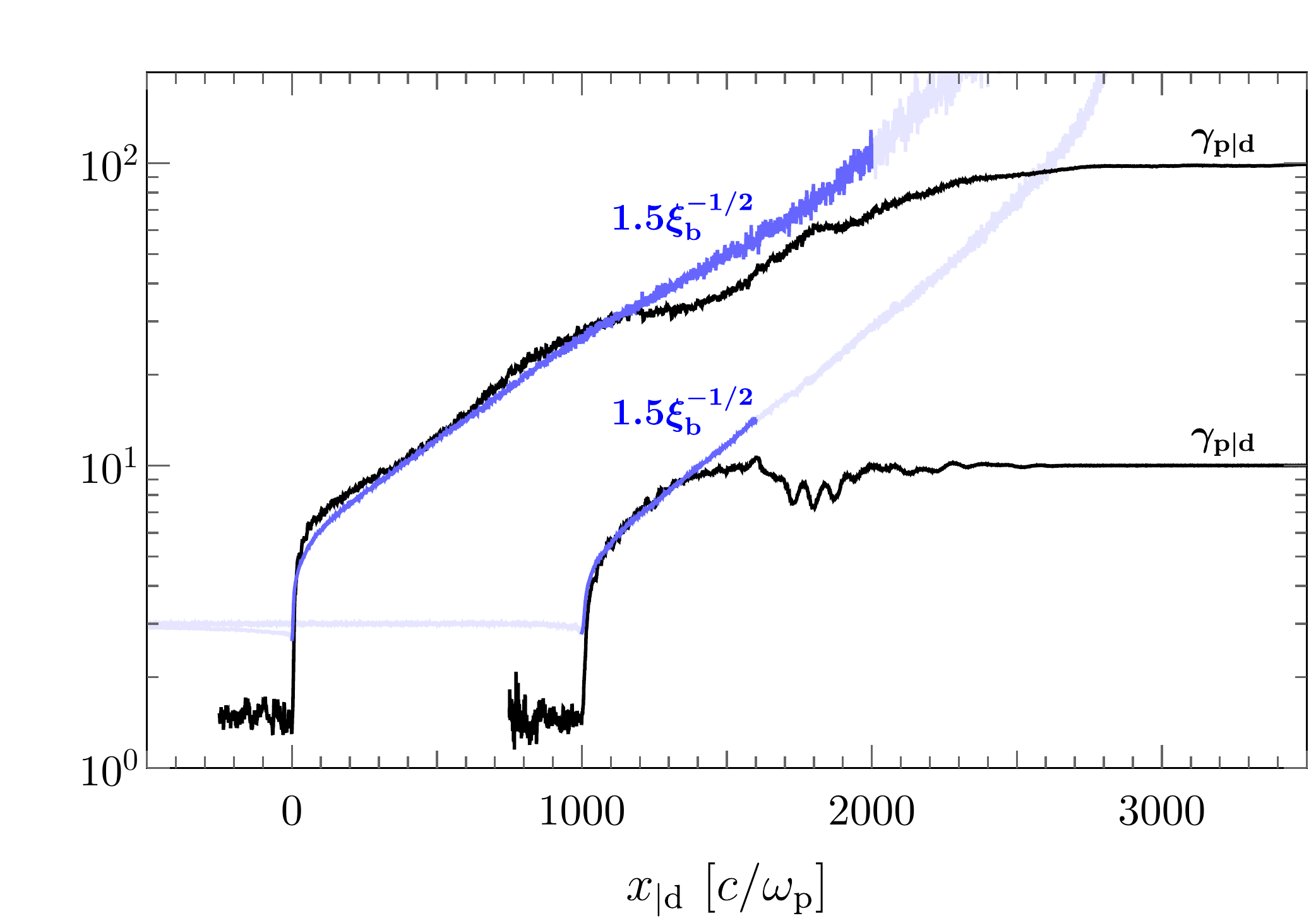}
 \caption{Spatial profiles of $\gamma_{\rm p\vert d}$ extracted from PIC simulations with $\gamma_{\infty\vert\rm d} = 100$ (top) and $\gamma_{\infty\vert\rm d} = 10$ (bottom) compared to the
 fluid deceleration law $ \sim \xi_{\rm b}^{-1/2}$ ($1.5$ an {\it ad hoc} factor); the data for $\gamma_{\infty\vert\rm d} = 10$ have been offset in $x_{\rm\vert d}$ by $1000 c/\omega_{\rm p}$ for clarity. 
  \label{fig:gp} }
\end{figure}

At the fluid level, this momentum transfer can be described via the scattering of suprathermal particles off an effectively magnetized background plasma. The ensuing deceleration of the background plasma can be quantified using the steady-state conservation laws for current and energy-momentum in $\mathcal R_{\rm s}$ \cite{L1_supp}. Through its interaction with the suprathermal beam, the background plasma loses an energy flux density $T_{\rm b}^{tx}$ and a momentum flux density $T_{\rm b}^{xx}$ between $+\infty$ and a point $x$ in the precursor. By definition of $\xi_{\rm b}$, one has $T_{\rm b}^{tx} \sim T_{\rm b}^{xx} \sim \xi_{\rm b} F_\infty$, up to different prefactors of the order of unity. In analogy with the
fact that, unless a particle moves at the same velocity $\beta_\infty$ as the plasma, its energy increases by $\gamma_\infty^2$ when it is picked up by the latter, one can show that deceleration occurs once $\vert \beta_\infty T_{\rm b}^{tx}-T_{\rm b}^{xx}\vert \gtrsim F_\infty/\gamma_\infty^2$~\cite{pap2}. Hence, where $\xi_{\rm b} \gtrsim 1/\gamma_\infty^2$, the Lorentz factor $\gamma_{\rm p}$ drops according to $\gamma_{\rm p}^2\xi_{\rm b} \simeq \rm const$. Figure~\ref{fig:gp} shows that this fluid deceleration law is well verified in PIC simulations.

Consequently, the background plasma slows down to subrelativistic velocities once $\xi_{\rm b} \sim 0.1-0.3$. This nicely accounts for the universal -- {\it i.e.}, independent of $\gamma_\infty$ -- fraction of shock energy injected into the suprathermal population, of typical value $\xi_{\rm b} \sim 0.1$ in PIC simulations, {\it e.g.}~\cite{2013ApJ...771...54S} and Fig.~\ref{fig:prof_ga100}.  Such cosmic-ray mediated shocks have been predicted in the subrelativistic limit~\cite{1981ApJ...248..344D} and observed in nonlinear Monte Carlo simulations~\cite{2016MNRAS.456.3090E}. One clear prediction is the existence of a sub-shock: in Fig.~\ref{fig:prof_ga100}, the Lorentz factor of the background plasma indeed decreases over thousands of $c/\omega_{\rm p}$ from $\gamma_{\infty\vert\rm d}$ down to $\gamma_{\rm sub} \sim 5$, at which point the shock transition suddenly occurs over $\lesssim 100 c/\omega_{\rm p}$, as discussed further below.

\noindent
\emph{The heating of the background plasma--}
The noninertial nature of the turbulence frame controls the heating of the background plasma as follows. In $\mathcal R_{\rm w}$, particles are subject to an effective gravity
$\propto {\rm d}u_{\rm w}/{\rm d}x$ directed toward the shock front, and to pitch-angle (elastic) scattering off the turbulence. This gives rise to Joule-like heating wherein gravity plays the role of the driving electric field, while turbulence-induced scattering provides collisional friction. The corresponding physics can be described by a general relativistic Vlasov-Fokker-Planck equation written in a mixed coordinate frame, with spatial variables in $\mathcal R_{\rm s}$ and momenta in $\mathcal R_{\rm w}$ \cite{L1_supp, pap2}. Here, we simulate this interplay between gravity and friction through a numerical Monte Carlo integration of the analog stochastic dynamical system: 
${\rm d}\mu_{\rm w} = \sqrt{2\nu_{\vert\rm w}{\rm d}t_{\vert\rm w}}\varsigma$ and ${\rm d}p_{\vert\rm w}^x = p_{\vert\rm w}{\rm d}\mu_{\rm w} - \left(\beta_{\rm w} {p_{\vert\rm w}^t}+{p_{\vert\rm w}^x}\right)
({\rm d} u_{\rm w}/{\rm d}x) {\rm d}t_{\vert\rm w}$, with $\mu_{\rm w} = p_{\rm \vert w}^x/p_{\rm \vert w}$  the pitch-angle cosine, $\nu_{\vert\rm w}$  the effective pitch-angle scattering frequency (treated as a constant parameter),  $\varsigma \sim \mathcal N(0,1)$ describing white noise and ${\rm d}t_{\rm \vert w}$  a time interval in $\mathcal R_{\rm w}$.

\begin{figure}
\includegraphics[width=0.45\textwidth]{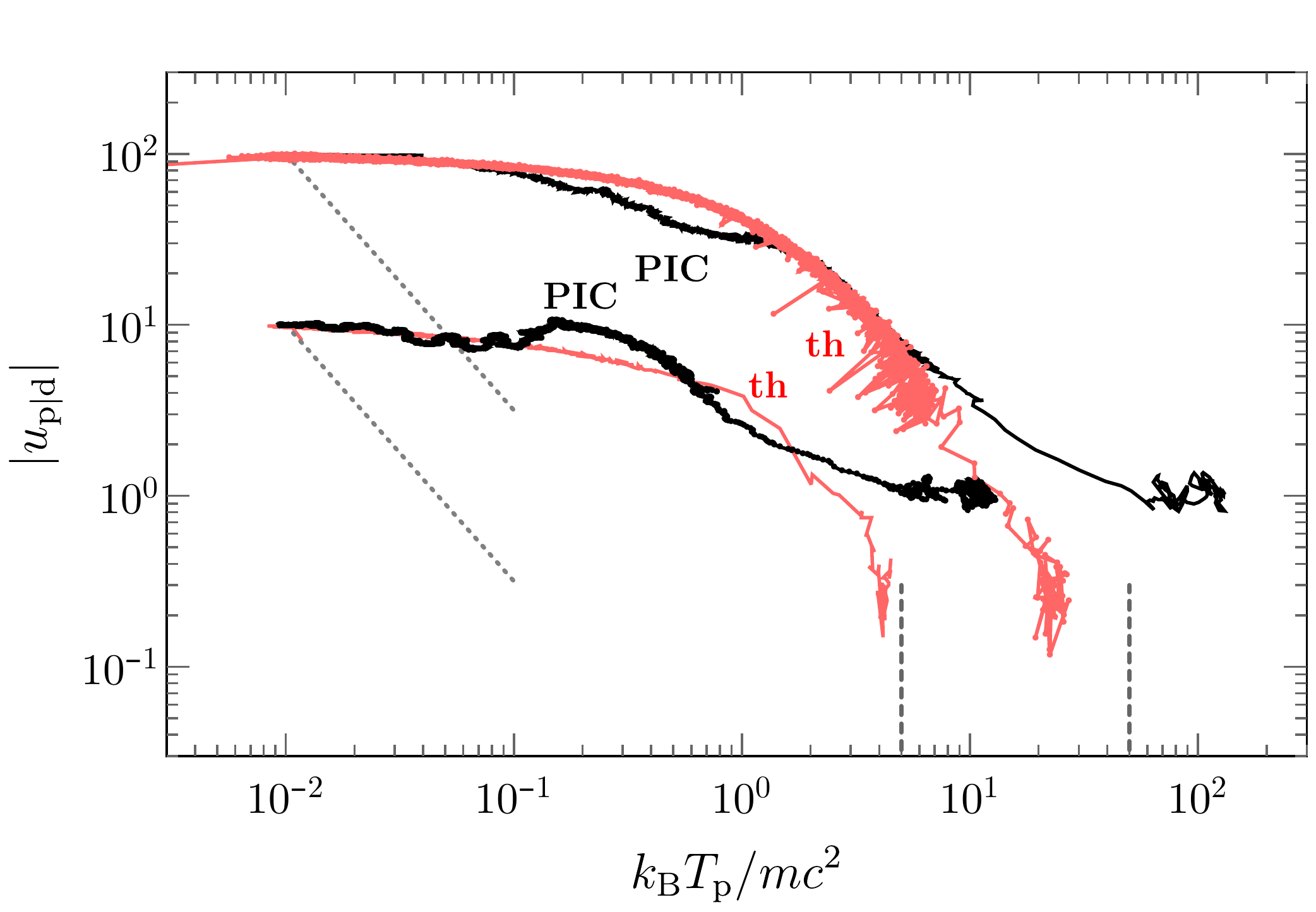}
 \caption{Trajectory of the background plasma in the temperature $T_p$ and $4$-velocity $\vert{u_{\rm p\vert d}}\vert$ plane, as measured in our reference PIC simulations (black) with $\gamma_{\infty\vert\rm d} = 100$ (top) and $\gamma_{\infty\vert\rm d} = 10$ (bottom), and as evaluated through numerical Monte Carlo integration for $\nu_{\vert\rm w} = 0.01\omega_{\rm p}$ (red). Dashed lines indicate the expected temperatures corresponding to the fluid shock jump conditions; dotted lines show the adiabatic compression law $T_{\rm p} \propto \vert {u_{\rm p}}\vert^{-2/3}$.
 \label{fig:upTp} }
\end{figure}

One can anticipate the trajectories ${u_{\rm p}}$ {\it vs} $T_{\rm p}$ along the plasma world line: for $\nu_{\vert\rm w} \rightarrow +\infty$, the background plasma behaves as a perfect fluid, hence $T_{\rm p} \propto \vert {u_{\rm p}}\vert^{-2/3}$, as befits adiabatic 1D compression of a subrelativistic fluid, while if $\nu_{\vert\rm w} \rightarrow 0$, the background plasma remains in
its initial state because it cannot experience the effective gravity.
Using our model (ii) for $u_{\rm w}(x)$ (Fig.~\ref{fig:bwp}), we obtain numerical predictions for $T_{\rm p}$ and $u_{\rm p}$ and compare them to the PIC simulation results in Fig.~\ref{fig:upTp}. For an effective $\nu_{\vert\rm w} = 0.01 \omega_{\rm p}$, the model trajectories satisfactorily reproduce those observed in the PIC simulations as well as the shock jump conditions.

Our PIC simulations reveal that about half of background plasma particles have experienced at least one turnaround while crossing the precursor. Given a precursor length
scale of $\ell_{\rm \vert d}\simeq 2\times 10^3 c/\omega_{\rm p}$  for $\gamma_{\infty\vert d}=100$ (resp. $\ell_{\rm\vert d}\simeq 10^3 c/\omega_{\rm p}$ for $\gamma_{\infty\vert d}=10$), boosting back to $\mathcal R_{\rm w}$ with typical Lorentz factor
$\langle\gamma_{\rm w\vert d}\rangle\sim 30$ over the precursor (resp. $\langle\gamma_{\rm w\vert d}\rangle\sim 10$), where this average is obtained as the value of $\gamma_{\rm p\vert d}$ halfway through the precursor, we estimate $\nu_{\vert\rm w} \sim \gamma_{\rm w\vert d} c/\ell_{\rm\vert d} \sim 1.5\times 10^{-2} \omega_{\rm p}$ (resp. $\sim 10^{-2} \omega_{\rm p}$),
which agrees with the value inferred above. Alternatively, assuming marginally untrapped particles, one expects $\nu_{\vert\rm w} \sim c r_\perp/r_{\rm g\vert w}^2$
($r_\perp \sim 1-10 c/\omega_{\rm p}$ the filament radius, $r_{\rm g\vert w}$ the gyroradius of a particle of Lorentz factor $\gamma_{\vert\rm w}$), {\it i.e.} $\nu_{\vert\rm w} \sim \epsilon_B\omega_{\rm p}/\gamma_{\rm\vert w}^2$, again consistent with the above value for mildly relativistic background plasma particles in $\mathcal R_{\rm w}$. In the case of bound particles oscillating
transversely in the filament at the betatron frequency $\omega_{\beta\vert\rm w} \sim c\left(r_{\perp}r_{\rm g\vert w}\right)^{-1/2}$, and experiencing decoherence of the force on a length
scale $r_{\parallel\vert\rm w}$, we derive $\nu_{\vert\rm w} \sim \omega_{\beta\vert\rm w}^2r_\perp^2/r_{\parallel\vert\rm w}c \sim \epsilon_B^{1/2} (r_\perp/r_{\parallel\vert\rm w}) \omega_{\rm p}/\gamma_{\rm\vert w}$, in fair agreement with the above results for a typical aspect ratio $r_\perp/r_{\parallel\vert\rm w} \sim 0.1$.

In the noninertial local plasma rest frame, heating thus occurs from $T_\infty \ll m_e c^2/k_{\rm B}$ up to $\sim \gamma_\infty m_e c^2/k_{\rm B}$ at the shock. In the shock frame, however, this dissipative
dynamics is better seen as the effect of a collisionless viscosity, which transfers momentum from the forward to the transverse directions while preserving the energy per particle, consistent
with the shock jump conditions~\cite{1976PhFl...19.1130B}. In this respect, electron heating in pair shocks starkly differs from that in electron-ion shocks: there, the electron population behaves as an open system that draws energy from the ion reservoir, so that electron energization truly occurs in $\mathcal{R}_{\rm s}$ through their interaction with transverse or longitudinal electromagnetic fields \cite{2006ApJ...641..978M, *2008PhRvE..77b6403G, *2012EL.....9735002G, *2013MNRAS.430.1280P, Kumar_2015}.

Although an electrostatic (longitudinal) electric field component indeed appears in our pair simulations, its contribution to the energization of particles in the near precursor ($x \lesssim 10^3 c/\omega_{\rm p}$)
is at best comparable to that of the transverse component, as we have checked using test particles. In the far precursor, the electrostatic contribution is significant, presumably due to the
excitation of non-transverse CFI modes, but the amount of heating there is small, see Fig.~\ref{fig:prof_ga100}. By contrast, the longitudinal electric field is expected to play an important role in electron-ion shocks~\cite{Kumar_2015} because the dependence of $\nu_{\vert\rm w}$ on inertia breaks the equivalence of the electron and ion trajectories in the effective gravity field. This might well preheat the electrons up to near equipartition with the ions, as observed numerically~\cite{2013ApJ...771...54S}. We further speculate that the above physics
of slowdown and heating in pair shocks could describe reasonably well the dynamics of the inertia-carrier ions in electron-ion collisionless shocks. Such study is left for further work.

\noindent
\emph{The shock transition--}
The relativistic motion of $\mathcal R_{\rm w}$ relative to $\mathcal R_{\rm s}$ also affects the growth rate of the CFI, which determines the profile of $\epsilon_B$: a background plasma element, subject to the CFI with a growth rate $\Im\omega_{\vert\rm w}$ in $\mathcal R_{\rm w}$ indeed experiences growth over a timescale $\tau = \gamma_{\rm w}/\Im\omega_{\vert\rm w}$ in
$\mathcal R_{\rm s}$. While the profile of $\epsilon_B(x)$ shows slow growth at large distances, its relatively flat shape at $x \lesssim 10^3 c/\omega_{\rm p}$
(Fig.~\ref{fig:prof_ga100}) is ascribed to saturation by advection of the CFI, {\it i.e.}, the $e-$folding scale is then larger than $x$. This is confirmed by detailed kinetic calculations of the growth rates \cite{pap4}.

\begin{figure}
\includegraphics[width=0.45\textwidth]{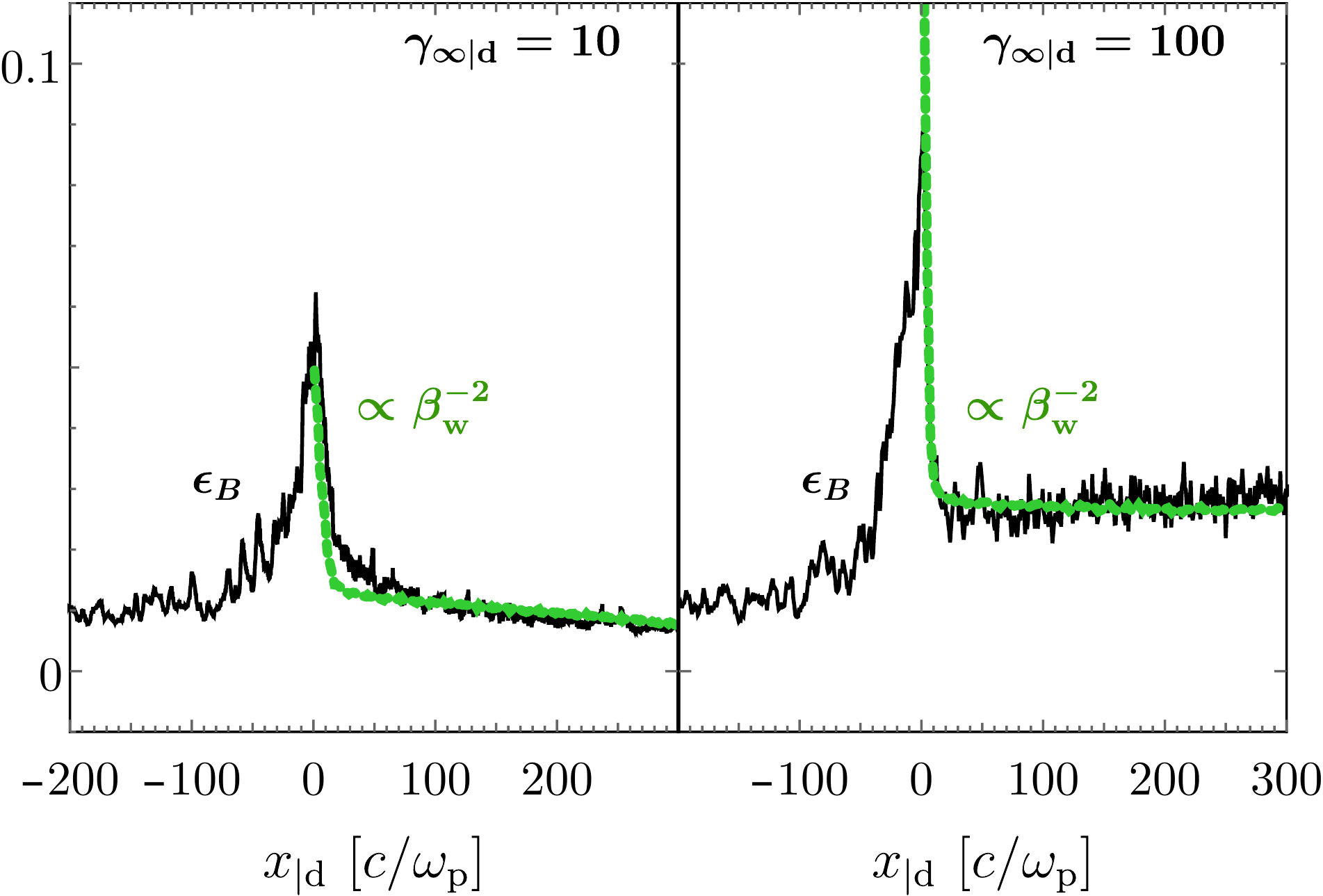}
 \caption{Zoom on the peak of $\epsilon_B$ at the shock, compared to the compression law $\epsilon_B \propto \beta_{\rm w}^{-2}$ (dashed green), where $\beta_{\rm w}$ is extracted from the
 PIC simulation through $\langle\delta E_y^2\rangle^{1/2}/\langle\delta B_z^2\rangle^{1/2}$, then smoothed. The proportionality factors are {\it ad hoc} linear fits to the scaling of $\epsilon_B$ at
 large $x \gtrsim 100 c/\omega_{\rm p}$. 
 \label{fig:epsB} }
\end{figure}

By contrast, the sudden rise in $\epsilon_B$ at the shock transition is commonly interpreted as the buildup of the electromagnetic barrier through microinstabilities. A nagging question is then: why does this occur precisely at the location predicted by the shock trajectory in the lab frame? Our model readily explains this sudden rise as the compression of transverse magnetic field lines in a decelerating flow: in $\mathcal R_{\rm s}$, the steady-state law $\boldsymbol{\nabla}\times\boldsymbol{\delta}\mathbf{E} = 0$ implies, for purely transverse fields, $\beta_{\rm w}\delta B_\perp={\rm const.}$, hence $\epsilon_B \propto \beta_{\rm w}^{-2}$. $\epsilon_B$ thus increases by $\sim10$ in the shock transition where $\beta_{\rm w}$ jumps from $\simeq -1$ to $\simeq -1/3$ ($-1/2$ in 2D), but remains constant elsewhere (up to instability growth). A detailed comparison of $\beta_{\rm w}(x)^{-2}$ and $\epsilon_B(x)$ nicely confirms the above, see Fig.~\ref{fig:epsB}.

The relativistic motion of $\mathcal R_{\rm w}$ relative to $\mathcal R_{\rm s}$ also impacts the scattering length scale $l_{\rm scatt}(p)$ of beam particles, which controls the acceleration
physics and the size of the shock transition (see below). In $\mathcal R_{\rm w}$, the standard estimate is $l_{\rm scatt\vert w}(p)\simeq r_{\rm g\vert w}^2/\lambda_{\delta B}$ for particles of
gyroradius $r_{\rm g\vert w}$ much larger than the coherence length $\lambda_{\delta B}\sim c/\omega_{\rm p}$. Converting it to $\mathcal R_{\rm s}$ brings in an additional prefactor $\gamma_{\rm w}$, $l_{\rm scatt}(p)\approx \gamma_{\rm w} r_{\rm g}^2\omega_{\rm p}/c = \gamma_{\rm w} \epsilon_B^{-1} (p/p_{\rm m})^2 c/\omega_{\rm p}$, with $p_{\rm m} = \gamma_\infty m_ec$ being the typical injection momentum. This formula is supported by a more elaborate quasilinear calculation that takes proper account of relativistic effects and of the anisotropic turbulent spectrum, and is found to match fairly well measurements from PIC simulations~\cite{pap3}. The $\gamma_{\rm w}$ factor, which results from the relativistic motion, implies a large acceleration timescale in the far precursor. This may explain why PIC simulations of limited duration observe Fermi acceleration mainly through grazing orbits on the shock front, where $\gamma_{\rm w} \sim 1$. 

Finally, at the shock transition, $\gamma_{\rm w} \sim 1$, $\epsilon_B \sim 0.1$ and $p \sim p_{\rm m}$, and hence $l_{\rm scatt}\sim 10-100 c/\omega_{\rm p}$. This  value, lower than that expected hundreds of $c/\omega_{\rm p}$ away, where $\epsilon_B \lesssim 0.01$ and $\gamma_{\rm w} \gg 1$, introduces a scattering barrier that selects the most energetic particles from the shocked thermal pool to form the population of injected suprathermal particles. The quadratic energy dependence of $l_{\rm scatt}$ implies that $\xi_{\rm b}(x)$ takes on a
powerlaw form, at least in the near precursor (Fig.~\ref{fig:prof_ga100}), which accounts for the powerlaw profile for $\gamma_{\rm p\vert d}$ (Fig.~\ref{fig:gp}). In this context, the core of the sub-shock can be interpreted as resulting from the pressure of suprathermal particles located within one scattering length from the shock front.
The background plasma decouples from the microturbulence once $\nu_{\rm\vert w}^{-1}$ exceeds the deceleration length scale of $\mathcal R_{\rm w}$, before it eventually relaxes in the asymptotic stationary $\mathcal R_{\rm w}$ frame, again on a $\sim \nu_{\rm\vert w}^{-1}$ length scale. Overall, we infer from Figs.~\ref{fig:bwp} and \ref{fig:gp} a sub-shock width of $\sim 100\,c/\omega_{\rm p}$, in reasonable agreement with the relaxation length ${\nu_{\vert\rm w}}^{-1}$.

In summary, we have presented an analytical microphysical model of the precursor of unmagnetized, relativistic collisionless pair shocks, which sheds new light into the phenomenology of such systems in high-energy and laboratory astrophysics. Our theory, based on the properties of the noninertial ``Weibel frame'' and benchmarked against large-scale PIC simulations, lends itself to extrapolation to other shock regimes and to the large spatiotemporal scales of astrophysical interest.

\noindent \emph{Acknowledgments:} We acknowledge financial support from the Programme National Hautes \'Energies (PNHE) of the C.N.R.S., the ANR-14-CE33-0019 MACH project and the ILP LABEX (reference ANR-10-LABX-63) as part of the Idex SUPER (reference ANR-11-IDEX-0004-02). This work was granted access to the HPC resources of TGCC/CCRT under the allocation 2018-A0030407666 made by GENCI. We also acknowledge PRACE for awarding us access to resource Joliot Curie-SKL at TGCC/CCRT.

\bibliographystyle{apsrev4-1}

\bibliography{shock}

\begin{thebibliography}{60}%
\makeatletter
\providecommand \@ifxundefined [1]{%
 \@ifx{#1\undefined}
}%
\providecommand \@ifnum [1]{%
 \ifnum #1\expandafter \@firstoftwo
 \else \expandafter \@secondoftwo
 \fi
}%
\providecommand \@ifx [1]{%
 \ifx #1\expandafter \@firstoftwo
 \else \expandafter \@secondoftwo
 \fi
}%
\providecommand \natexlab [1]{#1}%
\providecommand \enquote  [1]{``#1''}%
\providecommand \bibnamefont  [1]{#1}%
\providecommand \bibfnamefont [1]{#1}%
\providecommand \citenamefont [1]{#1}%
\providecommand \href@noop [0]{\@secondoftwo}%
\providecommand \href [0]{\begingroup \@sanitize@url \@href}%
\providecommand \@href[1]{\@@startlink{#1}\@@href}%
\providecommand \@@href[1]{\endgroup#1\@@endlink}%
\providecommand \@sanitize@url [0]{\catcode `\\12\catcode `\$12\catcode
  `\&12\catcode `\#12\catcode `\^12\catcode `\_12\catcode `\%12\relax}%
\providecommand \@@startlink[1]{}%
\providecommand \@@endlink[0]{}%
\providecommand \url  [0]{\begingroup\@sanitize@url \@url }%
\providecommand \@url [1]{\endgroup\@href {#1}{\urlprefix }}%
\providecommand \urlprefix  [0]{URL }%
\providecommand \Eprint [0]{\href }%
\providecommand \doibase [0]{http://dx.doi.org/}%
\providecommand \selectlanguage [0]{\@gobble}%
\providecommand \bibinfo  [0]{\@secondoftwo}%
\providecommand \bibfield  [0]{\@secondoftwo}%
\providecommand \translation [1]{[#1]}%
\providecommand \BibitemOpen [0]{}%
\providecommand \bibitemStop [0]{}%
\providecommand \bibitemNoStop [0]{.\EOS\space}%
\providecommand \EOS [0]{\spacefactor3000\relax}%
\providecommand \BibitemShut  [1]{\csname bibitem#1\endcsname}%
\let\auto@bib@innerbib\@empty
\bibitem [{\citenamefont {{Treumann}}(2009)}]{2009A&ARv..17..409T}%
  \BibitemOpen
  \bibfield  {author} {\bibinfo {author} {\bibfnamefont {R.~A.}\ \bibnamefont
  {{Treumann}}},\ }\href {\doibase 10.1007/s00159-009-0024-2} {\bibfield
  {journal} {\bibinfo  {journal} {Astron. Astrophys. Rev.}\ }\textbf {\bibinfo
  {volume} {17}},\ \bibinfo {pages} {409} (\bibinfo {year} {2009})}\BibitemShut
  {NoStop}%
\bibitem [{\citenamefont {{Bykov}}\ and\ \citenamefont
  {{Treumann}}(2011)}]{2011A&ARv..19...42B}%
  \BibitemOpen
  \bibfield  {author} {\bibinfo {author} {\bibfnamefont {A.~M.}\ \bibnamefont
  {{Bykov}}}\ and\ \bibinfo {author} {\bibfnamefont {R.~A.}\ \bibnamefont
  {{Treumann}}},\ }\href {\doibase 10.1007/s00159-011-0042-8} {\bibfield
  {journal} {\bibinfo  {journal} {Astron. Astrophys. Rev.}\ }\textbf {\bibinfo
  {volume} {19}},\ \bibinfo {eid} {42} (\bibinfo {year} {2011})}\BibitemShut
  {NoStop}%
\bibitem [{\citenamefont {{Marcowith}}\ \emph {et~al.}(2016)\citenamefont
  {{Marcowith}}, \citenamefont {{Bret}}, \citenamefont {{Bykov}}, \citenamefont
  {{Dieckman}}, \citenamefont {{O'C Drury}}, \citenamefont {{Lemb{\`e}ge}},
  \citenamefont {{Lemoine}}, \citenamefont {{Morlino}}, \citenamefont
  {{Murphy}}, \citenamefont {{Pelletier}}, \citenamefont {{Plotnikov}},
  \citenamefont {{Reville}}, \citenamefont {{Riquelme}}, \citenamefont
  {{Sironi}},\ and\ \citenamefont {{Stockem Novo}}}]{2016RPPh...79d6901M}%
  \BibitemOpen
  \bibfield  {author} {\bibinfo {author} {\bibfnamefont {A.}~\bibnamefont
  {{Marcowith}}}, \bibinfo {author} {\bibfnamefont {A.}~\bibnamefont {{Bret}}},
  \bibinfo {author} {\bibfnamefont {A.}~\bibnamefont {{Bykov}}}, \bibinfo
  {author} {\bibfnamefont {M.~E.}\ \bibnamefont {{Dieckman}}}, \bibinfo
  {author} {\bibfnamefont {L.}~\bibnamefont {{O'C Drury}}}, \bibinfo {author}
  {\bibfnamefont {B.}~\bibnamefont {{Lemb{\`e}ge}}}, \bibinfo {author}
  {\bibfnamefont {M.}~\bibnamefont {{Lemoine}}}, \bibinfo {author}
  {\bibfnamefont {G.}~\bibnamefont {{Morlino}}}, \bibinfo {author}
  {\bibfnamefont {G.}~\bibnamefont {{Murphy}}}, \bibinfo {author}
  {\bibfnamefont {G.}~\bibnamefont {{Pelletier}}}, \bibinfo {author}
  {\bibfnamefont {I.}~\bibnamefont {{Plotnikov}}}, \bibinfo {author}
  {\bibfnamefont {B.}~\bibnamefont {{Reville}}}, \bibinfo {author}
  {\bibfnamefont {M.}~\bibnamefont {{Riquelme}}}, \bibinfo {author}
  {\bibfnamefont {L.}~\bibnamefont {{Sironi}}}, \ and\ \bibinfo {author}
  {\bibfnamefont {A.}~\bibnamefont {{Stockem Novo}}},\ }\href {\doibase
  10.1088/0034-4885/79/4/046901} {\bibfield  {journal} {\bibinfo  {journal}
  {Rep. Prog. Phys.}\ }\textbf {\bibinfo {volume} {79}},\ \bibinfo {eid}
  {046901} (\bibinfo {year} {2016})}\BibitemShut {NoStop}%
\bibitem [{\citenamefont {{Blandford}}\ and\ \citenamefont
  {{Eichler}}(1987)}]{1987PhR...154....1B}%
  \BibitemOpen
  \bibfield  {author} {\bibinfo {author} {\bibfnamefont {R.}~\bibnamefont
  {{Blandford}}}\ and\ \bibinfo {author} {\bibfnamefont {D.}~\bibnamefont
  {{Eichler}}},\ }\href {\doibase 10.1016/0370-1573(87)90134-7} {\bibfield
  {journal} {\bibinfo  {journal} {Phys. Rep.}\ }\textbf {\bibinfo {volume}
  {154}},\ \bibinfo {pages} {1} (\bibinfo {year} {1987})}\BibitemShut {NoStop}%
\bibitem [{\citenamefont {{Abbott et
  al.}}(2017{\natexlab{a}})}]{PhysRevLett.119.161101}%
  \BibitemOpen
  \bibfield  {author} {\bibinfo {author} {\bibfnamefont {B.~P.}\ \bibnamefont
  {{Abbott et al.}}} (\bibinfo {collaboration} {LIGO Scientific Collaboration
  and Virgo Collaboration}),\ }\href {\doibase 10.1103/PhysRevLett.119.161101}
  {\bibfield  {journal} {\bibinfo  {journal} {Phys. Rev. Lett.}\ }\textbf
  {\bibinfo {volume} {119}},\ \bibinfo {pages} {161101} (\bibinfo {year}
  {2017}{\natexlab{a}})}\BibitemShut {NoStop}%
\bibitem [{\citenamefont {{Abbott et
  al.}}(2017{\natexlab{b}})}]{2041-8205-848-2-L12}%
  \BibitemOpen
  \bibfield  {author} {\bibinfo {author} {\bibfnamefont {B.~P.}\ \bibnamefont
  {{Abbott et al.}}},\ }\href {http://stacks.iop.org/2041-8205/848/i=2/a=L12}
  {\bibfield  {journal} {\bibinfo  {journal} {Astrophys. J.}\ }\textbf
  {\bibinfo {volume} {848}},\ \bibinfo {pages} {L12} (\bibinfo {year}
  {2017}{\natexlab{b}})}\BibitemShut {NoStop}%
\bibitem [{\citenamefont {Fox}\ \emph {et~al.}(2013)\citenamefont {Fox},
  \citenamefont {Fiksel}, \citenamefont {Bhattacharjee}, \citenamefont {Chang},
  \citenamefont {Germaschewski}, \citenamefont {Hu},\ and\ \citenamefont
  {Nilson}}]{PhysRevLett.111.225002}%
  \BibitemOpen
  \bibfield  {author} {\bibinfo {author} {\bibfnamefont {W.}~\bibnamefont
  {Fox}}, \bibinfo {author} {\bibfnamefont {G.}~\bibnamefont {Fiksel}},
  \bibinfo {author} {\bibfnamefont {A.}~\bibnamefont {Bhattacharjee}}, \bibinfo
  {author} {\bibfnamefont {P.-Y.}\ \bibnamefont {Chang}}, \bibinfo {author}
  {\bibfnamefont {K.}~\bibnamefont {Germaschewski}}, \bibinfo {author}
  {\bibfnamefont {S.~X.}\ \bibnamefont {Hu}}, \ and\ \bibinfo {author}
  {\bibfnamefont {P.~M.}\ \bibnamefont {Nilson}},\ }\href {\doibase
  10.1103/PhysRevLett.111.225002} {\bibfield  {journal} {\bibinfo  {journal}
  {Phys. Rev. Lett.}\ }\textbf {\bibinfo {volume} {111}},\ \bibinfo {pages}
  {225002} (\bibinfo {year} {2013})}\BibitemShut {NoStop}%
\bibitem [{\citenamefont {{Huntington et
  al.}}(2015)}]{Huntington_NP_11_173_2015}%
  \BibitemOpen
  \bibfield  {author} {\bibinfo {author} {\bibfnamefont {C.~M.}\ \bibnamefont
  {{Huntington et al.}}},\ }\href {\doibase 10.1038/nphys3178} {\bibfield
  {journal} {\bibinfo  {journal} {Nat. Phys.}\ }\textbf {\bibinfo {volume}
  {11}},\ \bibinfo {pages} {173} (\bibinfo {year} {2015})}\BibitemShut
  {NoStop}%
\bibitem [{\citenamefont {Ross~et al.}(2017)}]{PhysRevLett.118.185003}%
  \BibitemOpen
  \bibfield  {author} {\bibinfo {author} {\bibfnamefont {J.~S.}\ \bibnamefont
  {Ross~et al.}},\ }\href {\doibase 10.1103/PhysRevLett.118.185003} {\bibfield
  {journal} {\bibinfo  {journal} {Phys. Rev. Lett.}\ }\textbf {\bibinfo
  {volume} {118}},\ \bibinfo {pages} {185003} (\bibinfo {year}
  {2017})}\BibitemShut {NoStop}%
\bibitem [{\citenamefont {{Moiseev}}\ and\ \citenamefont
  {{Sagdeev}}(1963)}]{1963JNuE....5...43M}%
  \BibitemOpen
  \bibfield  {author} {\bibinfo {author} {\bibfnamefont {S.~S.}\ \bibnamefont
  {{Moiseev}}}\ and\ \bibinfo {author} {\bibfnamefont {R.~Z.}\ \bibnamefont
  {{Sagdeev}}},\ }\href {\doibase 10.1088/0368-3281/5/1/309} {\bibfield
  {journal} {\bibinfo  {journal} {J. Nuc. Energy}\ }\textbf {\bibinfo {volume}
  {5}},\ \bibinfo {pages} {43} (\bibinfo {year} {1963})}\BibitemShut {NoStop}%
\bibitem [{\citenamefont {{Medvedev}}\ and\ \citenamefont
  {{Loeb}}(1999)}]{1999ApJ...526..697M}%
  \BibitemOpen
  \bibfield  {author} {\bibinfo {author} {\bibfnamefont {M.~V.}\ \bibnamefont
  {{Medvedev}}}\ and\ \bibinfo {author} {\bibfnamefont {A.}~\bibnamefont
  {{Loeb}}},\ }\href {\doibase 10.1086/308038} {\bibfield  {journal} {\bibinfo
  {journal} {Astrophys. J.}\ }\textbf {\bibinfo {volume} {526}},\ \bibinfo
  {pages} {697} (\bibinfo {year} {1999})}\BibitemShut {NoStop}%
\bibitem [{\citenamefont {{Lyubarsky}}\ and\ \citenamefont
  {{Eichler}}(2006)}]{2006ApJ...647.1250L}%
  \BibitemOpen
  \bibfield  {author} {\bibinfo {author} {\bibfnamefont {Y.}~\bibnamefont
  {{Lyubarsky}}}\ and\ \bibinfo {author} {\bibfnamefont {D.}~\bibnamefont
  {{Eichler}}},\ }\href {\doibase 10.1086/505523} {\bibfield  {journal}
  {\bibinfo  {journal} {Astrophys. J.}\ }\textbf {\bibinfo {volume} {647}},\
  \bibinfo {pages} {1250} (\bibinfo {year} {2006})}\BibitemShut {NoStop}%
\bibitem [{\citenamefont {{Achterberg}}\ and\ \citenamefont
  {{Wiersma}}(2007)}]{2007A&A...475....1A}%
  \BibitemOpen
  \bibfield  {author} {\bibinfo {author} {\bibfnamefont {A.}~\bibnamefont
  {{Achterberg}}}\ and\ \bibinfo {author} {\bibfnamefont {J.}~\bibnamefont
  {{Wiersma}}},\ }\href {\doibase 10.1051/0004-6361:20065365} {\bibfield
  {journal} {\bibinfo  {journal} {Astron. Astrophys.}\ }\textbf {\bibinfo
  {volume} {475}},\ \bibinfo {pages} {1} (\bibinfo {year} {2007})}\BibitemShut
  {NoStop}%
\bibitem [{\citenamefont {{Achterberg}}\ \emph {et~al.}(2007)\citenamefont
  {{Achterberg}}, \citenamefont {{Wiersma}},\ and\ \citenamefont
  {{Norman}}}]{2007A&A...475...19A}%
  \BibitemOpen
  \bibfield  {author} {\bibinfo {author} {\bibfnamefont {A.}~\bibnamefont
  {{Achterberg}}}, \bibinfo {author} {\bibfnamefont {J.}~\bibnamefont
  {{Wiersma}}}, \ and\ \bibinfo {author} {\bibfnamefont {C.~A.}\ \bibnamefont
  {{Norman}}},\ }\href {\doibase 10.1051/0004-6361:20065366} {\bibfield
  {journal} {\bibinfo  {journal} {Astron. Astrophys.}\ }\textbf {\bibinfo
  {volume} {475}},\ \bibinfo {pages} {19} (\bibinfo {year} {2007})}\BibitemShut
  {NoStop}%
\bibitem [{\citenamefont {{Kato}}(2007)}]{2007ApJ...668..974K}%
  \BibitemOpen
  \bibfield  {author} {\bibinfo {author} {\bibfnamefont {T.~N.}\ \bibnamefont
  {{Kato}}},\ }\href {\doibase 10.1086/521297} {\bibfield  {journal} {\bibinfo
  {journal} {Astrophys. J.}\ }\textbf {\bibinfo {volume} {668}},\ \bibinfo
  {pages} {974} (\bibinfo {year} {2007})}\BibitemShut {NoStop}%
\bibitem [{\citenamefont
  {{Spitkovsky}}(2008{\natexlab{a}})}]{2008ApJ...673L..39S}%
  \BibitemOpen
  \bibfield  {author} {\bibinfo {author} {\bibfnamefont {A.}~\bibnamefont
  {{Spitkovsky}}},\ }\href {\doibase 10.1086/527374} {\bibfield  {journal}
  {\bibinfo  {journal} {Astrophys. J. Lett.}\ }\textbf {\bibinfo {volume}
  {673}},\ \bibinfo {pages} {L39} (\bibinfo {year}
  {2008}{\natexlab{a}})}\BibitemShut {NoStop}%
\bibitem [{\citenamefont {{Kato}}\ and\ \citenamefont
  {{Takabe}}(2008)}]{2008ApJ...681L..93K}%
  \BibitemOpen
  \bibfield  {author} {\bibinfo {author} {\bibfnamefont {T.~N.}\ \bibnamefont
  {{Kato}}}\ and\ \bibinfo {author} {\bibfnamefont {H.}~\bibnamefont
  {{Takabe}}},\ }\href {\doibase 10.1086/590387} {\bibfield  {journal}
  {\bibinfo  {journal} {Astrophys. J. Lett.}\ }\textbf {\bibinfo {volume}
  {681}},\ \bibinfo {pages} {L93} (\bibinfo {year} {2008})}\BibitemShut
  {NoStop}%
\bibitem [{\citenamefont
  {{Spitkovsky}}(2008{\natexlab{b}})}]{2008ApJ...682L...5S}%
  \BibitemOpen
  \bibfield  {author} {\bibinfo {author} {\bibfnamefont {A.}~\bibnamefont
  {{Spitkovsky}}},\ }\href {\doibase 10.1086/590248} {\bibfield  {journal}
  {\bibinfo  {journal} {Astrophys. J. Lett.}\ }\textbf {\bibinfo {volume}
  {682}},\ \bibinfo {pages} {L5} (\bibinfo {year}
  {2008}{\natexlab{b}})}\BibitemShut {NoStop}%
\bibitem [{\citenamefont {{Nishikawa}}\ \emph {et~al.}(2009)\citenamefont
  {{Nishikawa}}, \citenamefont {{Niemiec}}, \citenamefont {{Hardee}},
  \citenamefont {{Medvedev}}, \citenamefont {{Sol}}, \citenamefont {{Mizuno}},
  \citenamefont {{Zhang}}, \citenamefont {{Pohl}}, \citenamefont {{Oka}},\ and\
  \citenamefont {{Hartmann}}}]{2009ApJ...698L..10N}%
  \BibitemOpen
  \bibfield  {author} {\bibinfo {author} {\bibfnamefont {K.-I.}\ \bibnamefont
  {{Nishikawa}}}, \bibinfo {author} {\bibfnamefont {J.}~\bibnamefont
  {{Niemiec}}}, \bibinfo {author} {\bibfnamefont {P.~E.}\ \bibnamefont
  {{Hardee}}}, \bibinfo {author} {\bibfnamefont {M.}~\bibnamefont
  {{Medvedev}}}, \bibinfo {author} {\bibfnamefont {H.}~\bibnamefont {{Sol}}},
  \bibinfo {author} {\bibfnamefont {Y.}~\bibnamefont {{Mizuno}}}, \bibinfo
  {author} {\bibfnamefont {B.}~\bibnamefont {{Zhang}}}, \bibinfo {author}
  {\bibfnamefont {M.}~\bibnamefont {{Pohl}}}, \bibinfo {author} {\bibfnamefont
  {M.}~\bibnamefont {{Oka}}}, \ and\ \bibinfo {author} {\bibfnamefont {D.~H.}\
  \bibnamefont {{Hartmann}}},\ }\href {\doibase 10.1088/0004-637X/698/1/L10}
  {\bibfield  {journal} {\bibinfo  {journal} {Astrophys. J. Lett.}\ }\textbf
  {\bibinfo {volume} {698}},\ \bibinfo {pages} {L10} (\bibinfo {year}
  {2009})}\BibitemShut {NoStop}%
\bibitem [{\citenamefont {{Martins}}\ \emph {et~al.}(2009)\citenamefont
  {{Martins}}, \citenamefont {{Fonseca}}, \citenamefont {{Silva}},\ and\
  \citenamefont {{Mori}}}]{2009ApJ...695L.189M}%
  \BibitemOpen
  \bibfield  {author} {\bibinfo {author} {\bibfnamefont {S.~F.}\ \bibnamefont
  {{Martins}}}, \bibinfo {author} {\bibfnamefont {R.~A.}\ \bibnamefont
  {{Fonseca}}}, \bibinfo {author} {\bibfnamefont {L.~O.}\ \bibnamefont
  {{Silva}}}, \ and\ \bibinfo {author} {\bibfnamefont {W.~B.}\ \bibnamefont
  {{Mori}}},\ }\href {\doibase 10.1088/0004-637X/695/2/L189} {\bibfield
  {journal} {\bibinfo  {journal} {Astrophys. J. Lett.}\ }\textbf {\bibinfo
  {volume} {695}},\ \bibinfo {pages} {L189} (\bibinfo {year}
  {2009})}\BibitemShut {NoStop}%
\bibitem [{\citenamefont {{Sironi}}\ and\ \citenamefont
  {{Spitkovsky}}(2011)}]{2011ApJ...726...75S}%
  \BibitemOpen
  \bibfield  {author} {\bibinfo {author} {\bibfnamefont {L.}~\bibnamefont
  {{Sironi}}}\ and\ \bibinfo {author} {\bibfnamefont {A.}~\bibnamefont
  {{Spitkovsky}}},\ }\href {\doibase 10.1088/0004-637X/726/2/75} {\bibfield
  {journal} {\bibinfo  {journal} {ApJ}\ }\textbf {\bibinfo {volume} {726}},\
  \bibinfo {eid} {75} (\bibinfo {year} {2011})}\BibitemShut {NoStop}%
\bibitem [{\citenamefont {{Haugb{\o}lle}}(2011)}]{2011ApJ...739L..42H}%
  \BibitemOpen
  \bibfield  {author} {\bibinfo {author} {\bibfnamefont {T.}~\bibnamefont
  {{Haugb{\o}lle}}},\ }\href {\doibase 10.1088/2041-8205/739/2/L42} {\bibfield
  {journal} {\bibinfo  {journal} {Astrophys. J. Lett.}\ }\textbf {\bibinfo
  {volume} {739}},\ \bibinfo {eid} {L42} (\bibinfo {year} {2011})}\BibitemShut
  {NoStop}%
\bibitem [{\citenamefont {{Bret}}\ \emph {et~al.}(2008)\citenamefont {{Bret}},
  \citenamefont {{Gremillet}}, \citenamefont {{B{\'e}nisti}},\ and\
  \citenamefont {{Lefebvre}}}]{2008PhRvL.100t5008B}%
  \BibitemOpen
  \bibfield  {author} {\bibinfo {author} {\bibfnamefont {A.}~\bibnamefont
  {{Bret}}}, \bibinfo {author} {\bibfnamefont {L.}~\bibnamefont {{Gremillet}}},
  \bibinfo {author} {\bibfnamefont {D.}~\bibnamefont {{B{\'e}nisti}}}, \ and\
  \bibinfo {author} {\bibfnamefont {E.}~\bibnamefont {{Lefebvre}}},\ }\href
  {\doibase 10.1103/PhysRevLett.100.205008} {\bibfield  {journal} {\bibinfo
  {journal} {Phys. Rev. Lett.}\ }\textbf {\bibinfo {volume} {100}},\ \bibinfo
  {eid} {205008} (\bibinfo {year} {2008})}\BibitemShut {NoStop}%
\bibitem [{\citenamefont {{Bret}}\ \emph
  {et~al.}(2010{\natexlab{a}})\citenamefont {{Bret}}, \citenamefont
  {{Gremillet}},\ and\ \citenamefont {{Dieckmann}}}]{2010PhPl...17l0501B}%
  \BibitemOpen
  \bibfield  {author} {\bibinfo {author} {\bibfnamefont {A.}~\bibnamefont
  {{Bret}}}, \bibinfo {author} {\bibfnamefont {L.}~\bibnamefont {{Gremillet}}},
  \ and\ \bibinfo {author} {\bibfnamefont {M.~E.}\ \bibnamefont
  {{Dieckmann}}},\ }\href {\doibase 10.1063/1.3514586} {\bibfield  {journal}
  {\bibinfo  {journal} {Phys. Plasmas}\ }\textbf {\bibinfo {volume} {17}},\
  \bibinfo {pages} {120501} (\bibinfo {year} {2010}{\natexlab{a}})}\BibitemShut
  {NoStop}%
\bibitem [{\citenamefont {{Bret}}\ \emph
  {et~al.}(2010{\natexlab{b}})\citenamefont {{Bret}}, \citenamefont
  {{Gremillet}},\ and\ \citenamefont {{B{\'e}nisti}}}]{2010PhRvE..81c6402B}%
  \BibitemOpen
  \bibfield  {author} {\bibinfo {author} {\bibfnamefont {A.}~\bibnamefont
  {{Bret}}}, \bibinfo {author} {\bibfnamefont {L.}~\bibnamefont {{Gremillet}}},
  \ and\ \bibinfo {author} {\bibfnamefont {D.}~\bibnamefont {{B{\'e}nisti}}},\
  }\href {\doibase 10.1103/PhysRevE.81.036402} {\bibfield  {journal} {\bibinfo
  {journal} {Phys. Rev. E}\ }\textbf {\bibinfo {volume} {81}},\ \bibinfo {eid}
  {036402} (\bibinfo {year} {2010}{\natexlab{b}})}\BibitemShut {NoStop}%
\bibitem [{\citenamefont {{Lemoine}}\ and\ \citenamefont
  {{Pelletier}}(2010)}]{2010MNRAS.402..321L}%
  \BibitemOpen
  \bibfield  {author} {\bibinfo {author} {\bibfnamefont {M.}~\bibnamefont
  {{Lemoine}}}\ and\ \bibinfo {author} {\bibfnamefont {G.}~\bibnamefont
  {{Pelletier}}},\ }\href {\doibase 10.1111/j.1365-2966.2009.15869.x}
  {\bibfield  {journal} {\bibinfo  {journal} {MNRAS}\ }\textbf {\bibinfo
  {volume} {402}},\ \bibinfo {pages} {321} (\bibinfo {year}
  {2010})}\BibitemShut {NoStop}%
\bibitem [{\citenamefont {{Rabinak}}\ \emph {et~al.}(2011)\citenamefont
  {{Rabinak}}, \citenamefont {{Katz}},\ and\ \citenamefont
  {{Waxman}}}]{2011ApJ...736..157R}%
  \BibitemOpen
  \bibfield  {author} {\bibinfo {author} {\bibfnamefont {I.}~\bibnamefont
  {{Rabinak}}}, \bibinfo {author} {\bibfnamefont {B.}~\bibnamefont {{Katz}}}, \
  and\ \bibinfo {author} {\bibfnamefont {E.}~\bibnamefont {{Waxman}}},\ }\href
  {\doibase 10.1088/0004-637X/736/2/157} {\bibfield  {journal} {\bibinfo
  {journal} {ApJ}\ }\textbf {\bibinfo {volume} {736}},\ \bibinfo {eid} {157}
  (\bibinfo {year} {2011})}\BibitemShut {NoStop}%
\bibitem [{\citenamefont {{Lemoine}}\ and\ \citenamefont
  {{Pelletier}}(2011)}]{2011MNRAS.417.1148L}%
  \BibitemOpen
  \bibfield  {author} {\bibinfo {author} {\bibfnamefont {M.}~\bibnamefont
  {{Lemoine}}}\ and\ \bibinfo {author} {\bibfnamefont {G.}~\bibnamefont
  {{Pelletier}}},\ }\href {\doibase 10.1111/j.1365-2966.2011.19331.x}
  {\bibfield  {journal} {\bibinfo  {journal} {MNRAS}\ }\textbf {\bibinfo
  {volume} {417}},\ \bibinfo {pages} {1148} (\bibinfo {year}
  {2011})}\BibitemShut {NoStop}%
\bibitem [{\citenamefont {{Shaisultanov}}\ \emph {et~al.}(2012)\citenamefont
  {{Shaisultanov}}, \citenamefont {{Lyubarsky}},\ and\ \citenamefont
  {{Eichler}}}]{2012ApJ...744..182S}%
  \BibitemOpen
  \bibfield  {author} {\bibinfo {author} {\bibfnamefont {R.}~\bibnamefont
  {{Shaisultanov}}}, \bibinfo {author} {\bibfnamefont {Y.}~\bibnamefont
  {{Lyubarsky}}}, \ and\ \bibinfo {author} {\bibfnamefont {D.}~\bibnamefont
  {{Eichler}}},\ }\href {\doibase 10.1088/0004-637X/744/2/182} {\bibfield
  {journal} {\bibinfo  {journal} {ApJ}\ }\textbf {\bibinfo {volume} {744}},\
  \bibinfo {eid} {182} (\bibinfo {year} {2012})}\BibitemShut {NoStop}%
\bibitem [{\citenamefont {{Bret}}\ \emph {et~al.}(2013)\citenamefont {{Bret}},
  \citenamefont {{Stockem}}, \citenamefont {{Fiuza}}, \citenamefont {{Ruyer}},
  \citenamefont {{Gremillet}}, \citenamefont {{Narayan}},\ and\ \citenamefont
  {{Silva}}}]{2013PhPl...20d2102B}%
  \BibitemOpen
  \bibfield  {author} {\bibinfo {author} {\bibfnamefont {A.}~\bibnamefont
  {{Bret}}}, \bibinfo {author} {\bibfnamefont {A.}~\bibnamefont {{Stockem}}},
  \bibinfo {author} {\bibfnamefont {F.}~\bibnamefont {{Fiuza}}}, \bibinfo
  {author} {\bibfnamefont {C.}~\bibnamefont {{Ruyer}}}, \bibinfo {author}
  {\bibfnamefont {L.}~\bibnamefont {{Gremillet}}}, \bibinfo {author}
  {\bibfnamefont {R.}~\bibnamefont {{Narayan}}}, \ and\ \bibinfo {author}
  {\bibfnamefont {L.~O.}\ \bibnamefont {{Silva}}},\ }\href {\doibase
  10.1063/1.4798541} {\bibfield  {journal} {\bibinfo  {journal} {Phys.
  Plasmas}\ }\textbf {\bibinfo {volume} {20}},\ \bibinfo {eid} {042102}
  (\bibinfo {year} {2013})}\BibitemShut {NoStop}%
\bibitem [{\citenamefont {{Bret}}\ \emph {et~al.}(2014)\citenamefont {{Bret}},
  \citenamefont {{Stockem}}, \citenamefont {{Narayan}},\ and\ \citenamefont
  {{Silva}}}]{2014PhPl...21g2301B}%
  \BibitemOpen
  \bibfield  {author} {\bibinfo {author} {\bibfnamefont {A.}~\bibnamefont
  {{Bret}}}, \bibinfo {author} {\bibfnamefont {A.}~\bibnamefont {{Stockem}}},
  \bibinfo {author} {\bibfnamefont {R.}~\bibnamefont {{Narayan}}}, \ and\
  \bibinfo {author} {\bibfnamefont {L.~O.}\ \bibnamefont {{Silva}}},\ }\href
  {\doibase 10.1063/1.4886121} {\bibfield  {journal} {\bibinfo  {journal}
  {Phys. Plasmas}\ }\textbf {\bibinfo {volume} {21}},\ \bibinfo {eid} {072301}
  (\bibinfo {year} {2014})}\BibitemShut {NoStop}%
\bibitem [{\citenamefont {{Katz}}\ \emph {et~al.}(2007)\citenamefont {{Katz}},
  \citenamefont {{Keshet}},\ and\ \citenamefont
  {{Waxman}}}]{2007ApJ...655..375K}%
  \BibitemOpen
  \bibfield  {author} {\bibinfo {author} {\bibfnamefont {B.}~\bibnamefont
  {{Katz}}}, \bibinfo {author} {\bibfnamefont {U.}~\bibnamefont {{Keshet}}}, \
  and\ \bibinfo {author} {\bibfnamefont {E.}~\bibnamefont {{Waxman}}},\ }\href
  {\doibase 10.1086/509115} {\bibfield  {journal} {\bibinfo  {journal}
  {Astrophys. J.}\ }\textbf {\bibinfo {volume} {655}},\ \bibinfo {pages} {375}
  (\bibinfo {year} {2007})}\BibitemShut {NoStop}%
\bibitem [{\citenamefont {{Medvedev}}\ and\ \citenamefont
  {{Zakutnyaya}}(2009)}]{2009ApJ...696.2269M}%
  \BibitemOpen
  \bibfield  {author} {\bibinfo {author} {\bibfnamefont {M.~V.}\ \bibnamefont
  {{Medvedev}}}\ and\ \bibinfo {author} {\bibfnamefont {O.~V.}\ \bibnamefont
  {{Zakutnyaya}}},\ }\href {\doibase 10.1088/0004-637X/696/2/2269} {\bibfield
  {journal} {\bibinfo  {journal} {Astrophys. J.}\ }\textbf {\bibinfo {volume}
  {696}},\ \bibinfo {pages} {2269} (\bibinfo {year} {2009})}\BibitemShut
  {NoStop}%
\bibitem [{\citenamefont {{Ruyer}}\ \emph {et~al.}(2016)\citenamefont
  {{Ruyer}}, \citenamefont {{Gremillet}}, \citenamefont {{Bonnaud}},\ and\
  \citenamefont {{Riconda}}}]{2016PhRvL.117f5001R}%
  \BibitemOpen
  \bibfield  {author} {\bibinfo {author} {\bibfnamefont {C.}~\bibnamefont
  {{Ruyer}}}, \bibinfo {author} {\bibfnamefont {L.}~\bibnamefont
  {{Gremillet}}}, \bibinfo {author} {\bibfnamefont {G.}~\bibnamefont
  {{Bonnaud}}}, \ and\ \bibinfo {author} {\bibfnamefont {C.}~\bibnamefont
  {{Riconda}}},\ }\href {\doibase 10.1103/PhysRevLett.117.065001} {\bibfield
  {journal} {\bibinfo  {journal} {\prl}\ }\textbf {\bibinfo {volume} {117}},\
  \bibinfo {eid} {065001} (\bibinfo {year} {2016})}\BibitemShut {NoStop}%
\bibitem [{\citenamefont {{Ruyer}}\ \emph {et~al.}(2017)\citenamefont
  {{Ruyer}}, \citenamefont {{Gremillet}}, \citenamefont {{Bonnaud}},\ and\
  \citenamefont {{Riconda}}}]{2017PhPl...24d1409R}%
  \BibitemOpen
  \bibfield  {author} {\bibinfo {author} {\bibfnamefont {C.}~\bibnamefont
  {{Ruyer}}}, \bibinfo {author} {\bibfnamefont {L.}~\bibnamefont
  {{Gremillet}}}, \bibinfo {author} {\bibfnamefont {G.}~\bibnamefont
  {{Bonnaud}}}, \ and\ \bibinfo {author} {\bibfnamefont {C.}~\bibnamefont
  {{Riconda}}},\ }\href {\doibase 10.1063/1.4979187} {\bibfield  {journal}
  {\bibinfo  {journal} {Phys. Plasmas}\ }\textbf {\bibinfo {volume} {24}},\
  \bibinfo {eid} {041409} (\bibinfo {year} {2017})}\BibitemShut {NoStop}%
\bibitem [{\citenamefont {{Lefebvre}}\ \emph {et~al.}(2003)\citenamefont
  {{Lefebvre}}, \citenamefont {{Cochet}}, \citenamefont {{Fritzler}},
  \citenamefont {{Malka}}, \citenamefont {{Al{\'e}onard}}, \citenamefont
  {{Chemin}}, \citenamefont {{Darbon}}, \citenamefont {{Disdier}},
  \citenamefont {{Faure}}, \citenamefont {{Fedotoff}}, \citenamefont
  {{Landoas}}, \citenamefont {{Malka}}, \citenamefont {{M{\'e}ot}},
  \citenamefont {{Morel}}, \citenamefont {{Rabec LeGloahec}}, \citenamefont
  {{Rouyer}}, \citenamefont {{Rubbelynck}}, \citenamefont {{Tikhonchuk}},
  \citenamefont {{Wrobel}}, \citenamefont {{Audebert}},\ and\ \citenamefont
  {{Rousseaux}}}]{2003NucFu..43..629L}%
  \BibitemOpen
  \bibfield  {author} {\bibinfo {author} {\bibfnamefont {E.}~\bibnamefont
  {{Lefebvre}}}, \bibinfo {author} {\bibfnamefont {N.}~\bibnamefont
  {{Cochet}}}, \bibinfo {author} {\bibfnamefont {S.}~\bibnamefont
  {{Fritzler}}}, \bibinfo {author} {\bibfnamefont {V.}~\bibnamefont {{Malka}}},
  \bibinfo {author} {\bibfnamefont {M.~M.}\ \bibnamefont {{Al{\'e}onard}}},
  \bibinfo {author} {\bibfnamefont {J.~F.}\ \bibnamefont {{Chemin}}}, \bibinfo
  {author} {\bibfnamefont {S.}~\bibnamefont {{Darbon}}}, \bibinfo {author}
  {\bibfnamefont {L.}~\bibnamefont {{Disdier}}}, \bibinfo {author}
  {\bibfnamefont {J.}~\bibnamefont {{Faure}}}, \bibinfo {author} {\bibfnamefont
  {A.}~\bibnamefont {{Fedotoff}}}, \bibinfo {author} {\bibfnamefont
  {O.}~\bibnamefont {{Landoas}}}, \bibinfo {author} {\bibfnamefont
  {G.}~\bibnamefont {{Malka}}}, \bibinfo {author} {\bibfnamefont
  {V.}~\bibnamefont {{M{\'e}ot}}}, \bibinfo {author} {\bibfnamefont
  {P.}~\bibnamefont {{Morel}}}, \bibinfo {author} {\bibfnamefont
  {M.}~\bibnamefont {{Rabec LeGloahec}}}, \bibinfo {author} {\bibfnamefont
  {A.}~\bibnamefont {{Rouyer}}}, \bibinfo {author} {\bibfnamefont
  {C.}~\bibnamefont {{Rubbelynck}}}, \bibinfo {author} {\bibfnamefont
  {V.}~\bibnamefont {{Tikhonchuk}}}, \bibinfo {author} {\bibfnamefont
  {R.}~\bibnamefont {{Wrobel}}}, \bibinfo {author} {\bibfnamefont
  {P.}~\bibnamefont {{Audebert}}}, \ and\ \bibinfo {author} {\bibfnamefont
  {C.}~\bibnamefont {{Rousseaux}}},\ }\href {\doibase
  10.1088/0029-5515/43/7/317} {\bibfield  {journal} {\bibinfo  {journal} {Nuc.
  Fus.}\ }\textbf {\bibinfo {volume} {43}},\ \bibinfo {pages} {629} (\bibinfo
  {year} {2003})}\BibitemShut {NoStop}%
\bibitem [{L1_()}]{L1_supp}%
  \BibitemOpen
  \href@noop {} {}\bibinfo {note} {See Supplemental Material at [URL] for a
  description of the numerical simulations, including
  Refs.~\cite{Godfrey_JCP_2013,Godfrey_JCP_2014a,*Godfrey_JCP_2014b,Vay_JCP_2011,1997ITMTT..45..991C,
  *2002ITAP...50.1185C, *karkkainen2006low}, and of the modeling of the
  background plasma dynamics, involving
  Ref.~\cite{1985ApJ...296..319W,*1989ApJ...340.1112W}.}\BibitemShut {Stop}%
\bibitem [{Note1()}]{Note1}%
  \BibitemOpen
  \bibinfo {note} {Quantities related to the background plasma (resp.
  suprathermal beam) are indexed with $_{\protect \rm p}$ (resp. $_{\protect
  \rm b}$). Also, subscripts $_{\delimiter "026A30C \protect \rm d}$
  ($_{\delimiter "026A30C \protect \rm p}$) refer to quantities measured in the
  simulation (downstream) rest frame (resp. the background plasma rest
  frame).}\BibitemShut {Stop}%
\bibitem [{\citenamefont {{Pelletier}}\ \emph {et~al.}(2019)\citenamefont
  {{Pelletier}}, \citenamefont {{Gremillet}}, \citenamefont {{Vanthieghem}},\
  and\ \citenamefont {{Lemoine}}}]{pap1}%
  \BibitemOpen
  \bibfield  {author} {\bibinfo {author} {\bibfnamefont {G.}~\bibnamefont
  {{Pelletier}}}, \bibinfo {author} {\bibfnamefont {L.}~\bibnamefont
  {{Gremillet}}}, \bibinfo {author} {\bibfnamefont {A.}~\bibnamefont
  {{Vanthieghem}}}, \ and\ \bibinfo {author} {\bibfnamefont {M.}~\bibnamefont
  {{Lemoine}}},\ }\href {\doibase 10.1103/PhysRevE.100.013205} {\bibfield
  {journal} {\bibinfo  {journal} {Phys. Rev. E}\ }\textbf {\bibinfo {volume}
  {100}},\ \bibinfo {pages} {013205} (\bibinfo {year} {2019})},\ \Eprint
  {http://arxiv.org/abs/1907.07750} {arXiv:1907.07750} \BibitemShut {NoStop}%
\bibitem [{\citenamefont {{Lemoine}}\ \emph
  {et~al.}(2019{\natexlab{a}})\citenamefont {{Lemoine}}, \citenamefont
  {{Vanthieghem}}, \citenamefont {{Pelletier}},\ and\ \citenamefont
  {{Gremillet}}}]{pap2}%
  \BibitemOpen
  \bibfield  {author} {\bibinfo {author} {\bibfnamefont {M.}~\bibnamefont
  {{Lemoine}}}, \bibinfo {author} {\bibfnamefont {A.}~\bibnamefont
  {{Vanthieghem}}}, \bibinfo {author} {\bibfnamefont {G.}~\bibnamefont
  {{Pelletier}}}, \ and\ \bibinfo {author} {\bibfnamefont {L.}~\bibnamefont
  {{Gremillet}}},\ }\href@noop {} {\bibfield  {journal} {\bibinfo  {journal}
  {Phys. Rev. E, submitted (Pap. II)}\ } (\bibinfo {year}
  {2019}{\natexlab{a}})},\ \Eprint {http://arxiv.org/abs/1907.08219}
  {arXiv:1907.08219 [astro-ph.HE]} \BibitemShut {NoStop}%
\bibitem [{\citenamefont {{Sironi}}\ \emph {et~al.}(2013)\citenamefont
  {{Sironi}}, \citenamefont {{Spitkovsky}},\ and\ \citenamefont
  {{Arons}}}]{2013ApJ...771...54S}%
  \BibitemOpen
  \bibfield  {author} {\bibinfo {author} {\bibfnamefont {L.}~\bibnamefont
  {{Sironi}}}, \bibinfo {author} {\bibfnamefont {A.}~\bibnamefont
  {{Spitkovsky}}}, \ and\ \bibinfo {author} {\bibfnamefont {J.}~\bibnamefont
  {{Arons}}},\ }\href {\doibase 10.1088/0004-637X/771/1/54} {\bibfield
  {journal} {\bibinfo  {journal} {ApJ}\ }\textbf {\bibinfo {volume} {771}},\
  \bibinfo {eid} {54} (\bibinfo {year} {2013})}\BibitemShut {NoStop}%
\bibitem [{\citenamefont {{Drury}}\ and\ \citenamefont
  {{Voelk}}(1981)}]{1981ApJ...248..344D}%
  \BibitemOpen
  \bibfield  {author} {\bibinfo {author} {\bibfnamefont {L.~O.}\ \bibnamefont
  {{Drury}}}\ and\ \bibinfo {author} {\bibfnamefont {J.~H.}\ \bibnamefont
  {{Voelk}}},\ }\href {\doibase 10.1086/159159} {\bibfield  {journal} {\bibinfo
   {journal} {Astrophys. J.}\ }\textbf {\bibinfo {volume} {248}},\ \bibinfo
  {pages} {344} (\bibinfo {year} {1981})}\BibitemShut {NoStop}%
\bibitem [{\citenamefont {{Ellison}}\ \emph {et~al.}(2016)\citenamefont
  {{Ellison}}, \citenamefont {{Warren}},\ and\ \citenamefont
  {{Bykov}}}]{2016MNRAS.456.3090E}%
  \BibitemOpen
  \bibfield  {author} {\bibinfo {author} {\bibfnamefont {D.~C.}\ \bibnamefont
  {{Ellison}}}, \bibinfo {author} {\bibfnamefont {D.~C.}\ \bibnamefont
  {{Warren}}}, \ and\ \bibinfo {author} {\bibfnamefont {A.~M.}\ \bibnamefont
  {{Bykov}}},\ }\href {\doibase 10.1093/mnras/stv2912} {\bibfield  {journal}
  {\bibinfo  {journal} {MNRAS}\ }\textbf {\bibinfo {volume} {456}},\ \bibinfo
  {pages} {3090} (\bibinfo {year} {2016})}\BibitemShut {NoStop}%
\bibitem [{\citenamefont {{Blandford}}\ and\ \citenamefont
  {{McKee}}(1976)}]{1976PhFl...19.1130B}%
  \BibitemOpen
  \bibfield  {author} {\bibinfo {author} {\bibfnamefont {R.~D.}\ \bibnamefont
  {{Blandford}}}\ and\ \bibinfo {author} {\bibfnamefont {C.~F.}\ \bibnamefont
  {{McKee}}},\ }\href {\doibase 10.1063/1.861619} {\bibfield  {journal}
  {\bibinfo  {journal} {Phys. Fl.}\ }\textbf {\bibinfo {volume} {19}},\
  \bibinfo {pages} {1130} (\bibinfo {year} {1976})}\BibitemShut {NoStop}%
\bibitem [{\citenamefont {{Milosavljevi{\'c}}}\ and\ \citenamefont
  {{Nakar}}(2006)}]{2006ApJ...641..978M}%
  \BibitemOpen
  \bibfield  {author} {\bibinfo {author} {\bibfnamefont {M.}~\bibnamefont
  {{Milosavljevi{\'c}}}}\ and\ \bibinfo {author} {\bibfnamefont
  {E.}~\bibnamefont {{Nakar}}},\ }\href {\doibase 10.1086/500654} {\bibfield
  {journal} {\bibinfo  {journal} {Astrophys. J.}\ }\textbf {\bibinfo {volume}
  {641}},\ \bibinfo {pages} {978} (\bibinfo {year} {2006})}\BibitemShut
  {NoStop}%
\bibitem [{\citenamefont {{Gedalin}}\ \emph {et~al.}(2008)\citenamefont
  {{Gedalin}}, \citenamefont {{Balikhin}},\ and\ \citenamefont
  {{Eichler}}}]{2008PhRvE..77b6403G}%
  \BibitemOpen
  \bibfield  {author} {\bibinfo {author} {\bibfnamefont {M.}~\bibnamefont
  {{Gedalin}}}, \bibinfo {author} {\bibfnamefont {M.~A.}\ \bibnamefont
  {{Balikhin}}}, \ and\ \bibinfo {author} {\bibfnamefont {D.}~\bibnamefont
  {{Eichler}}},\ }\href {\doibase 10.1103/PhysRevE.77.026403} {\bibfield
  {journal} {\bibinfo  {journal} {Phys. Rev. E}\ }\textbf {\bibinfo {volume}
  {77}},\ \bibinfo {eid} {026403} (\bibinfo {year} {2008})}\BibitemShut
  {NoStop}%
\bibitem [{\citenamefont {{Gedalin}}\ \emph {et~al.}(2012)\citenamefont
  {{Gedalin}}, \citenamefont {{Smolik}}, \citenamefont {{Spitkovsky}},\ and\
  \citenamefont {{Balikhin}}}]{2012EL.....9735002G}%
  \BibitemOpen
  \bibfield  {author} {\bibinfo {author} {\bibfnamefont {M.}~\bibnamefont
  {{Gedalin}}}, \bibinfo {author} {\bibfnamefont {E.}~\bibnamefont {{Smolik}}},
  \bibinfo {author} {\bibfnamefont {A.}~\bibnamefont {{Spitkovsky}}}, \ and\
  \bibinfo {author} {\bibfnamefont {M.}~\bibnamefont {{Balikhin}}},\ }\href
  {\doibase 10.1209/0295-5075/97/35002} {\bibfield  {journal} {\bibinfo
  {journal} {Europhys. Lett.}\ }\textbf {\bibinfo {volume} {97}},\ \bibinfo
  {pages} {35002} (\bibinfo {year} {2012})}\BibitemShut {NoStop}%
\bibitem [{\citenamefont {{Plotnikov}}\ \emph {et~al.}(2013)\citenamefont
  {{Plotnikov}}, \citenamefont {{Pelletier}},\ and\ \citenamefont
  {{Lemoine}}}]{2013MNRAS.430.1280P}%
  \BibitemOpen
  \bibfield  {author} {\bibinfo {author} {\bibfnamefont {I.}~\bibnamefont
  {{Plotnikov}}}, \bibinfo {author} {\bibfnamefont {G.}~\bibnamefont
  {{Pelletier}}}, \ and\ \bibinfo {author} {\bibfnamefont {M.}~\bibnamefont
  {{Lemoine}}},\ }\href {\doibase 10.1093/mnras/sts696} {\bibfield  {journal}
  {\bibinfo  {journal} {MNRAS}\ }\textbf {\bibinfo {volume} {430}},\ \bibinfo
  {pages} {1280} (\bibinfo {year} {2013})}\BibitemShut {NoStop}%
\bibitem [{\citenamefont {{Kumar}}\ \emph {et~al.}(2015)\citenamefont
  {{Kumar}}, \citenamefont {{Eichler}},\ and\ \citenamefont
  {{Gedalin}}}]{Kumar_2015}%
  \BibitemOpen
  \bibfield  {author} {\bibinfo {author} {\bibfnamefont {R.}~\bibnamefont
  {{Kumar}}}, \bibinfo {author} {\bibfnamefont {D.}~\bibnamefont {{Eichler}}},
  \ and\ \bibinfo {author} {\bibfnamefont {M.}~\bibnamefont {{Gedalin}}},\
  }\href {\doibase 10.1088/0004-637X/806/2/165} {\bibfield  {journal} {\bibinfo
   {journal} {Astrophys. J.}\ }\textbf {\bibinfo {volume} {806}},\ \bibinfo
  {eid} {165} (\bibinfo {year} {2015})}\BibitemShut {NoStop}%
\bibitem [{\citenamefont {{Vanthieghem}}\ \emph {et~al.}(2019)\citenamefont
  {{Vanthieghem}}, \citenamefont {{Lemoine}}, \citenamefont {{Gremillet}},\
  and\ \citenamefont {{Pelletier}}}]{pap4}%
  \BibitemOpen
  \bibfield  {author} {\bibinfo {author} {\bibfnamefont {A.}~\bibnamefont
  {{Vanthieghem}}}, \bibinfo {author} {\bibfnamefont {M.}~\bibnamefont
  {{Lemoine}}}, \bibinfo {author} {\bibfnamefont {L.}~\bibnamefont
  {{Gremillet}}}, \ and\ \bibinfo {author} {\bibfnamefont {G.}~\bibnamefont
  {{Pelletier}}},\ }\href@noop {} {\bibfield  {journal} {\bibinfo  {journal}
  {Phys. Rev. E, in prep. (Pap. IV)}\ } (\bibinfo {year} {2019})}\BibitemShut
  {NoStop}%
\bibitem [{\citenamefont {{Lemoine}}\ \emph
  {et~al.}(2019{\natexlab{b}})\citenamefont {{Lemoine}}, \citenamefont
  {{Pelletier}}, \citenamefont {{Vanthieghem}},\ and\ \citenamefont
  {{Gremillet}}}]{pap3}%
  \BibitemOpen
  \bibfield  {author} {\bibinfo {author} {\bibfnamefont {M.}~\bibnamefont
  {{Lemoine}}}, \bibinfo {author} {\bibfnamefont {G.}~\bibnamefont
  {{Pelletier}}}, \bibinfo {author} {\bibfnamefont {A.}~\bibnamefont
  {{Vanthieghem}}}, \ and\ \bibinfo {author} {\bibfnamefont {L.}~\bibnamefont
  {{Gremillet}}},\ }\href@noop {} {\bibfield  {journal} {\bibinfo  {journal}
  {Phys. Rev. E, submitted (Pap. III)}\ } (\bibinfo {year}
  {2019}{\natexlab{b}})},\ \Eprint {http://arxiv.org/abs/1907.10294}
  {arXiv:1907.10294 [astro-ph.HE]} \BibitemShut {NoStop}%
\bibitem [{\citenamefont {{Godfrey}}\ and\ \citenamefont
  {{Vay}}(2013)}]{Godfrey_JCP_2013}%
  \BibitemOpen
  \bibfield  {author} {\bibinfo {author} {\bibfnamefont {B.~B.}\ \bibnamefont
  {{Godfrey}}}\ and\ \bibinfo {author} {\bibfnamefont {J.-L.}\ \bibnamefont
  {{Vay}}},\ }\href {\doibase 10.1016/j.jcp.2013.04.006} {\bibfield  {journal}
  {\bibinfo  {journal} {J. Comp. Phys.}\ }\textbf {\bibinfo {volume} {248}},\
  \bibinfo {pages} {33} (\bibinfo {year} {2013})}\BibitemShut {NoStop}%
\bibitem [{\citenamefont {{Godfrey}}\ \emph {et~al.}(2014)\citenamefont
  {{Godfrey}}, \citenamefont {{Vay}},\ and\ \citenamefont
  {{Haber}}}]{Godfrey_JCP_2014a}%
  \BibitemOpen
  \bibfield  {author} {\bibinfo {author} {\bibfnamefont {B.~B.}\ \bibnamefont
  {{Godfrey}}}, \bibinfo {author} {\bibfnamefont {J.-L.}\ \bibnamefont
  {{Vay}}}, \ and\ \bibinfo {author} {\bibfnamefont {I.}~\bibnamefont
  {{Haber}}},\ }\href {\doibase 10.1016/j.jcp.2013.10.053} {\bibfield
  {journal} {\bibinfo  {journal} {J. Comp. Phys.}\ }\textbf {\bibinfo {volume}
  {258}},\ \bibinfo {pages} {689} (\bibinfo {year} {2014})}\BibitemShut
  {NoStop}%
\bibitem [{\citenamefont {{Godfrey}}\ and\ \citenamefont
  {{Vay}}(2014)}]{Godfrey_JCP_2014b}%
  \BibitemOpen
  \bibfield  {author} {\bibinfo {author} {\bibfnamefont {B.~B.}\ \bibnamefont
  {{Godfrey}}}\ and\ \bibinfo {author} {\bibfnamefont {J.-L.}\ \bibnamefont
  {{Vay}}},\ }\href {\doibase 10.1016/j.jcp.2014.02.022} {\bibfield  {journal}
  {\bibinfo  {journal} {J. Comp. Phys.}\ }\textbf {\bibinfo {volume} {267}},\
  \bibinfo {pages} {1} (\bibinfo {year} {2014})}\BibitemShut {NoStop}%
\bibitem [{\citenamefont {{Vay}}\ \emph {et~al.}(2011)\citenamefont {{Vay}},
  \citenamefont {{Geddes}}, \citenamefont {{Cormier-Michel}},\ and\
  \citenamefont {{Grote}}}]{Vay_JCP_2011}%
  \BibitemOpen
  \bibfield  {author} {\bibinfo {author} {\bibfnamefont {J.-L.}\ \bibnamefont
  {{Vay}}}, \bibinfo {author} {\bibfnamefont {C.~G.~R.}\ \bibnamefont
  {{Geddes}}}, \bibinfo {author} {\bibfnamefont {E.}~\bibnamefont
  {{Cormier-Michel}}}, \ and\ \bibinfo {author} {\bibfnamefont {D.~P.}\
  \bibnamefont {{Grote}}},\ }\href {\doibase 10.1016/j.jcp.2011.04.003}
  {\bibfield  {journal} {\bibinfo  {journal} {J. Comp. Phys.}\ }\textbf
  {\bibinfo {volume} {230}},\ \bibinfo {pages} {5908} (\bibinfo {year}
  {2011})}\BibitemShut {NoStop}%
\bibitem [{\citenamefont {{Cole}}(1997)}]{1997ITMTT..45..991C}%
  \BibitemOpen
  \bibfield  {author} {\bibinfo {author} {\bibfnamefont {J.~B.}\ \bibnamefont
  {{Cole}}},\ }\href {\doibase 10.1109/22.588615} {\bibfield  {journal}
  {\bibinfo  {journal} {IEEE Trans. Microwave Theory Techniques}\ }\textbf
  {\bibinfo {volume} {45}},\ \bibinfo {pages} {991} (\bibinfo {year}
  {1997})}\BibitemShut {NoStop}%
\bibitem [{\citenamefont {{Cole}}(2002)}]{2002ITAP...50.1185C}%
  \BibitemOpen
  \bibfield  {author} {\bibinfo {author} {\bibfnamefont {J.~B.}\ \bibnamefont
  {{Cole}}},\ }\href {\doibase 10.1109/TAP.2002.801268} {\bibfield  {journal}
  {\bibinfo  {journal} {IEEE Trans. Antennas and Propagation}\ }\textbf
  {\bibinfo {volume} {50}},\ \bibinfo {pages} {1185} (\bibinfo {year}
  {2002})}\BibitemShut {NoStop}%
\bibitem [{\citenamefont {{K\"arkk\"ainen}}\ \emph {et~al.}(2006)\citenamefont
  {{K\"arkk\"ainen}}, \citenamefont {{Gjonaj}}, \citenamefont {{Lau}},\ and\
  \citenamefont {{Weiland}}}]{karkkainen2006low}%
  \BibitemOpen
  \bibfield  {author} {\bibinfo {author} {\bibfnamefont {M.}~\bibnamefont
  {{K\"arkk\"ainen}}}, \bibinfo {author} {\bibfnamefont {E.}~\bibnamefont
  {{Gjonaj}}}, \bibinfo {author} {\bibfnamefont {T.}~\bibnamefont {{Lau}}}, \
  and\ \bibinfo {author} {\bibfnamefont {T.}~\bibnamefont {{Weiland}}},\ }in\
  \href@noop {} {\emph {\bibinfo {booktitle} {Proc. Int. Comp. Acc. Phys.
  Conf., Chamonix, France}}}\ (\bibinfo {year} {2006})\ pp.\ \bibinfo {pages}
  {35--40}\BibitemShut {NoStop}%
\bibitem [{\citenamefont {{Webb}}(1985)}]{1985ApJ...296..319W}%
  \BibitemOpen
  \bibfield  {author} {\bibinfo {author} {\bibfnamefont {G.~M.}\ \bibnamefont
  {{Webb}}},\ }\href {\doibase 10.1086/163451} {\bibfield  {journal} {\bibinfo
  {journal} {Astrophys. J.}\ }\textbf {\bibinfo {volume} {296}},\ \bibinfo
  {pages} {319} (\bibinfo {year} {1985})}\BibitemShut {NoStop}%
\bibitem [{\citenamefont {{Webb}}(1989)}]{1989ApJ...340.1112W}%
  \BibitemOpen
  \bibfield  {author} {\bibinfo {author} {\bibfnamefont {G.~M.}\ \bibnamefont
  {{Webb}}},\ }\href {\doibase 10.1086/167462} {\bibfield  {journal} {\bibinfo
  {journal} {Astrophys. J.}\ }\textbf {\bibinfo {volume} {340}},\ \bibinfo
  {pages} {1112} (\bibinfo {year} {1989})}\BibitemShut {NoStop}%
\end{thebibliography}%

\end{document}